\documentclass[sigconf]{acmart}
\renewcommand\footnotetextcopyrightpermission[1]{} 

\newenvironment{sproof}{%
  \renewcommand{\proofname}{Proof Sketch}\proof}{\endproof}

\newcommand{\hide}[1]{}
\newcommand{\yb}[1]{{\color{black}#1}}
\newcommand{\rv}[1]{{\color{black}#1}}

\newcommand*\cim[1]{\tag*{\footnotesize\textit{(#1)}}}
\newcommand{\xhdr}[1]{\vspace{2mm}{\noindent\bfseries #1}.}
\usepackage{relsize}
\usepackage{subcaption}
\usepackage{tikz}
\usetikzlibrary{topaths,calc}
\usepackage[font=small]{caption}
\usepackage{amsmath}
\usepackage{enumitem}
\usepackage{csquotes}
\usepackage{algorithm}
\usepackage{xspace}
\usepackage{arydshln}
\usepackage{bbm}
\usepackage{algpseudocode}
\usepackage{amsthm}
\usepackage{etoolbox}
\usepackage{setspace}
\usepackage{wasysym}
\newcommand*{\eoc}{\ensuremath{\mathbb{E}^{c}}}
\newcommand{\method}{SHyRe\xspace}
\newcommand{\update}{\textbf{U{\footnotesize PDATE}}\xspace}
\newtheorem{defn}{Definition}
\AfterEndEnvironment{defn}{\noindent\ignorespaces}

\DeclareMathOperator*{\argmax}{argmax} 

\AtBeginDocument{%
  \providecommand\BibTeX{{%
    \normalfont B\kern-0.5em{\scshape i\kern-0.25em b}\kern-0.8em\TeX}}}

\setcopyright{none}
\begin{document}

\title{Supervised Hypergraph Reconstruction}

\author{Yanbang Wang}
\email{yw786@cornell.edu}
\affiliation{%
  \institution{Cornell University}
  \city{Ithaca}
  \state{New York}
  \country{USA}
}
\author{Jon Kleinberg}
\email{kleinberg@cornell.edu}
\affiliation{%
  \institution{Cornell University}
  \city{Ithaca}
  \state{New York}
  \country{USA}
}

\renewcommand{\shortauthors}{Wang and Kleinberg, et al.}

\begin{abstract}
We study an issue commonly seen with graph data analysis: many real-world complex systems involving high-order interactions are best encoded by hypergraphs; however, their datasets often end up being published or studied only in the form of their projections (with dyadic edges). To understand this issue, we first establish a theoretical framework to characterize this issue's implications and worst-case scenarios. The analysis motivates our formulation of the new task, supervised hypergraph reconstruction: reconstructing a real-world hypergraph from its projected graph, with the help of some existing knowledge of the application domain.

To reconstruct hypergraph data, we start by analyzing hyperedge distributions in the projection, based on which we create a framework containing two modules: (1) to handle the enormous search space of potential hyperedges, we design a sampling strategy with efficacy guarantees that significantly narrows the space to a smaller set of candidates; (2) to identify hyperedges from the candidates, we further design a hyperedge classifier in two well-working variants that capture structural features in the projection. Extensive experiments validate our claims, approach, and extensions. Remarkably, our approach outperforms all baselines by an order of magnitude in accuracy on hard datasets.
Our code and data can be downloaded from \url{bit.ly/SHyRe}.

\end{abstract}

\vspace{-1mm}
\vspace{-1mm}

\maketitle
\pagestyle{plain}
\section{Introduction}\label{sec:intro}

Graphs are a mathematical formalism that can describe many real-world complex systems by recording which pairs of entities in the system are connected, using the language of nodes and edges. Hypergraphs take this idea further by extending the concept of edges from pairwise relations to sets of arbitrary sizes, and thus admit a more expressive form of encoding for multilateral relationships. 

The long-standing problem that this paper addresses is that many datasets that should have been encoded by hypergraphs ended up being released or studied almost exclusively in graph format. There are many examples of this phenomenon: co-authorship networks \cite{newman2004coauthorship,sarigol2014predicting} where an edge encodes two author's collaboration in the same paper, social interaction networks \cite{madan2011sensing,klimt2004introducing} where an edge encodes two persons' interaction in a conversation or email, and protein-protein interaction networks \cite{safari2014protein} where an edge encodes two proteins' co-occurrence in one biological process. Co-authorships, conversations, and bio-processes all typically involve multiple authors, people, and proteins. 

Clearly, pairwise relations (\textit{i.e.} graphs) contain less information than the original hypergraphs they come from. Replacing hypergraphs with graphs is known to bias how people perceive \cite{wolf2016advantages}, predict \cite{arya2020hypersage}, and exploit \cite{arya2018exploiting} real-world interconnected systems. Despite the drawbacks, there are many real-world scenarios where the crucial underlying hypergraphs are dropped and only projected graphs get studied and released. These scenarios come in two cases:

\begin{itemize}[leftmargin=3mm]\vspace{-0.8mm}
    \item \rv{\textbf{Unobservable:} In some key scenarios, the available technology for data collection can only detect pairwise relations. This is most common in social science. For example, in \cite{madan2011sensing,ozella2021using,dai2020temporal}, sensing methodologies to record physical proximity can be used to build networks of face-to-face interaction: an interaction between two people is recorded by thresholding distances between their body-worn sensors. There is no way to directly record the multiple participants in each conversation by sensors only. Recovering multi-party events in a purely decentralized, peer-to-peer network system of communication is an important application.}\vspace{1mm}
    \item \rv{\textbf{Unpublished:} Even when they are technically observable, in practice the source hypergraph datasets of many studies are never released or published. For example, many of the most impactful studies analyzing coauthorships do not make available a hypergraph version of their dataset \cite{newman2004coauthorship,sarigol2014predicting}. Many popular graph learning benchmarks, including arXiv-hepth \cite{leskovec2005graphs}, ogbl-collab \cite{hu2020open}, and ogbn-product \cite{hu2020open}, also do not provide their hypergraph originals. Yet the underlying, unobserved hypergraphs contain important information about the domain. }
\end{itemize}\vspace{-1.2mm}

\rv{There do exist measures to recover a hypergraph, or at least a part of it, from graphs in these domains. Traditionally, such recovery process involves laborious manual work, and only one hypergraph can be recovered at a time, \textit{i.e.} the recovery is not generalizable to similar hypergraphs, and is hard to carry out at scale. To recover unobservable social interactions, social scientists use surveys that ask students to recall all people they had talked to in each daily conversation \cite{ozella2021using}; to obtain unpublished hypergraphs, researchers go great lengths to either 
 replicate tedious data preprocessing or to trace back to data publishers. Measures like these require a considerable amount of time, effort, and luck. 
}

\hide{Dyadic edges also have significant redundancy when used to encode high-order relations. Therefore as a side benefit a well-reconstructed hypergraph is a much more efficient way of storing high-order relations in large quantities.}


\begin{figure}[t]\vspace{-3mm}
    \centering
    \includegraphics[scale=0.54]{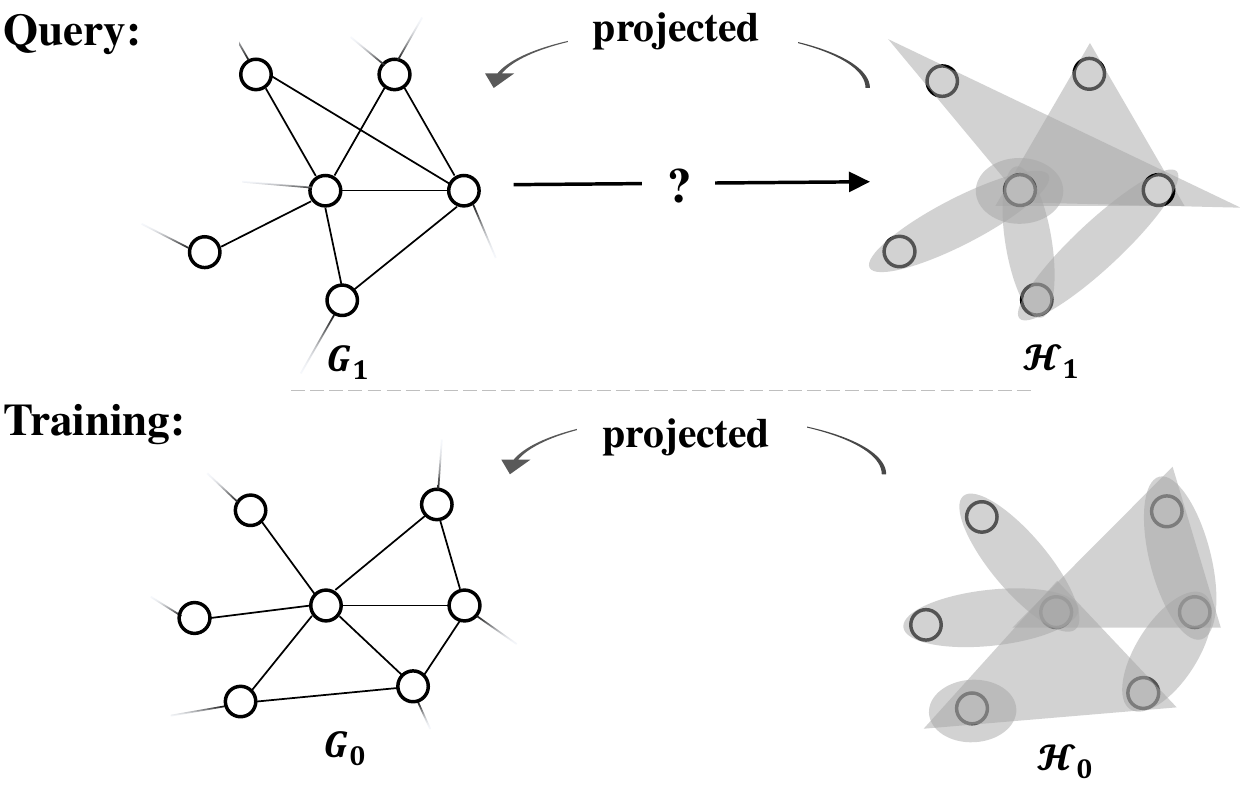}\vspace{-3mm}
    \caption{The supervised hypergraph reconstruction task. $\mathcal{H}_0$ and $\mathcal{H}_1$ belong to the same application domain. Given $\mathcal{H}_0$ (and its projection $G_0$), can we reconstruct $\mathcal{H}_1$ from its projection $G_1$?}
    \label{fig:task}
    \vspace{-4mm}
\end{figure}
All the scenarios above share a common problem abstraction, as follows.
There is an underlying hypergraph $\mathcal{H}_1$ that we cannot directly observe, and instead we only have access to its {\em projected graph} (or, {\em projection}) $G_1$, which we define to be the graph on the same node set as the hypergraph, and with two nodes connected in $G_1$ if and only if they belong to a shared hyperedge in $\mathcal{H}_1$.
Our goal is to reconstruct $\mathcal{H}_1$ as accurately as possible, given knowledge of $G_1$.
The ``query'' half in Fig.\ref{fig:task} illustrates this problem: 

We will later discuss a training phase where a known hypergraph from the same domain is provided. \yb{For broadest applicability, we focus on a hard version of the problem by assuming that the edges of the projected graph $G_1$ are unweighted; they just say whether two nodes appear in at least one hyperedge together.
In Appendix \ref{subsec:multiedge} we will also discuss an easier version of the problem where $G_1$ has weighted edges, recording the {\em number} of different hyperedges that each pair of nodes belongs to.}

{\em Our goal towards the reconstruction is two-fold:} \textbf{(1)} the reconstructed hypergraph can benefit data analysis and prediction modeling by providing efficiently-stored higher-order information for convenient usage; \textbf{(2)} the reconstruction process necessitates deeper understanding of hyperedge structure, especially (i) how hyperedges overlap each other in real-world datasets, and (ii) how different overlap patterns may obscure different amounts of high-order information once projected. Such understandings can help data contributors make more informed choice when they consider trade-offs in projecting higher-order structure out of their data.

\vspace{0.6mm}
\noindent\textbf{The challenge.} The biggest challenge in hypergraph reconstruction is that a given projected graph will gnenerally admit an enormous amount of possible hypergraphs. In simple cases, a hypergraph can be easily reconstructed if most of its hyperedges are disjoint --- simply by treating every maximal clique as a hyperedge. However, this can rarely be the case with real-world datasets, in which multiple hyperedges overlap each other, making some especially hard to detect. In Sec. \ref{sec:theory}, we formalize these difficult patterns of overlap and quantify their theoretical implications. We will also see that these patterns are ubiquitous in real-world datasets. Therefore, it is extremely hard to reconstruct a hypergraph arising from an arbitrary application domain without knowing anything about what hypergraphs look like in that domain.

\vspace{0.6mm}
\noindent\textbf{Previous work.} Due to the technical challenge, very little work was done on this problem, despite its significant value for applications. In graph mining, the closest tasks are hyperedge prediction and community detection.  Hyperedge prediction  \cite{Yadati2020NHP,xu2013hyperlink,benson2018simplicialbenson2018sequences} is a fundamentally different problem, in that the input to it is a hypergraph rather than its projected graph.  Additionally, methods for this problem typically only identify hyperedges from a given set of candidates, rather than the large implicit spaces of all possible hyperedges. Community detection, on the other hand, is based on looking for densely connected subsets of a graph under various definitions, but not for hyperedges. Both tasks are very different from our goal of searching hyperedges over the projection.

Besides the above, \cite{young2021hypergraph} is by far the closest related work. However, it aims to use least number of hyperedges to cover an arbitrary graph (see its Fig.7), as a way to explain the ``local density, global sparsity'' of some networks. Their work attempts parsimonious explanations of data rather than recovery of ground truth. Consequently, their ``principle of parsimony'' does not give high accuracy on many real-world hypergraphs, where there are significant hyperedge overlaps. We will see in Sec. \ref{sec:exp} that all these methods above yield significantly lower performance than our approach.

\vspace{0.6mm}
\rv{\noindent\textbf{Using supervised signal.} To help reconstruction in the presence of these forms of uncertainty, we propose the usage of training data: another hypergraph from the same application domain. Shown in Fig.\ref{fig:task}, our effort to reconstruct hypergraph $\mathcal{H}_1$ now involves two inputs: $\mathcal{H}_1$'s projection $G_1$, and an additional hypergraph $\mathcal{H}_0$ (along with its projection $G_0$) from the same or similar distribution as $\mathcal{H}_1$. 

The training data $\mathcal{H}_0$ is not only necessary but also reasonable to have. In practice, $\mathcal{H}_0$ often refers to manually collected data (as described earlier in this section), including surveys of participants (for social science applications), neuron groups labeled by experts for another common ancestral species or homologous organ (for biological or neuroscience applications), or a similar hypergraph obtained from another data publisher --- e.g.~for coauthorships, papers from a different database or of a different year slice.



For now, we assume $\mathcal{H}_0$ contains a comparable number of hyperedges to $\mathcal{H}_1$. In Sec. \ref{subsec:exp_semi} we generalize this to a semi-supervised setting where \rv{$\mathcal{H}_0$ is downsampled to a smaller fraction for training, and a transfer setting where $\mathcal{H}_0$ comes from a different distribution than $\mathcal{H}_1$.} As is standard in supervised learning, we consider it realistic to assume some knowledge about our reconstruction goal when searching through the enormous space of candidates.}

That said, inductiveness is enforced in our setting, meaning that two nodes with the same \rv{node ID} in $\mathcal{H}_0$ and $\mathcal{H}_1$ are irrelevant entities. It also means that \rv{we are not satisfied with solving one particular test case but also care about generalizability. More formally, using $\mathcal{H}_0$, we should be able to reconstruct not only $\mathcal{H}_1$ but also other hypergraphs from the same domain as $\mathcal{H}_1$.} \hide{$\mathcal{H}_2,\mathcal{H}_3, ...$ from the same domain.} Therefore, simply memorizing $\mathcal{H}_0$'s hyperedges does not work. 

\rv{With the help of training data, hypergraph reconstruction, which was previously almost impossible, now becomes more feasible in principle. Nevertheless, it still retains considerable challenges that we must address in designing solutions.}\hide{With the training data, the reconstruction now becomes a more feasible goal, though it still remains very challenging.}

\vspace{1mm}
\noindent\textbf{Our approach.} We start from the fact that every clique in the projection qualifies as a hyperedge. There are two critical steps to accomplish in order to get hyperedges from the cliques: 1. we cannot afford to traverse all cliques (since there can be too many), so a shortlisting procedure is necessary to significantly shrink the pool while retaining as many hyperedges as possible; 2. after obtaining the shortlist, we need to go through cliques in the shortlist and figure out which ones are real hyperedges.

It is extremely hard to accomplish any step above from the projected graph alone, so we leverage the supervised signal, \textit{$\mathcal{H}_0$}:
\begin{itemize}[leftmargin=3mm]\vspace{-0.8mm}
    \item For shortlisting, we design a sampling strategy based on a key observation of the consistency in hyperedge distribution between $\mathcal{H}_0$ and $\mathcal{H}_1$. We then optimize parameters of our strategy on $\mathcal{H}_0$, and use the optimized strategy to downsample the cliques of  $\mathcal{H}_1$.
    \item For identifying hyperedges, we design two variants of a hyperedge classifier. We train the classifier on cliques of $\mathcal{H}_0$, and then use it to identify hyperedges from cliques of $\mathcal{H}_1$. 
\end{itemize}

\vspace{-0.8mm}
\noindent\textbf{Contributions.} Our main contribution is three-fold:\vspace{-0.8mm}
\begin{itemize}[leftmargin=3mm]
    \item We identify the recovery of higher-order structure as an important$\hspace{0.5mm}$and pervasive issue commonly seen in graph data analysis. To understand the issue, we establish a topological analysis framework to characterize its implications and worst-case scenarios. The analysis motivates our formulation of a new task: supervised hypergraph reconstruction.
    \item We observe important structural properties of hyperedge distributions from projected graphs. Based on our observations, we design a new reconstruction framework, which contains a sampling strategy with theoretical guarantees, and a hyperedge classifier in two well-working variants.
    \item We conduct extensive experiments that validate our claims, approach, and extensions. We adapt 7 baselines and compare them with our approach on 8 real-world datasets. Our approach outperforms all baselines, and by orders of magnitude on hard datasets.
\end{itemize}

\hide{
\vspace{-0.5mm}
\noindent The rest of the paper is arranged as follows. Sec. \ref{sec:prelim} introduces the preliminaries, and Sec. \ref{sec:def} formalizes the problem. Sec. \ref{sec:theory} establishes a framework to analyze the projected graph and reconstruction from a theoretical perspective. Insights from Sec. \ref{sec:theory} underpin our framework, which is elaborated in Sec. \ref{sec:framework}. Sec. \ref{sec:exp} reports experiments validating our claims, approaches, and their extensions. }






\vspace{-1mm}
\section{Preliminaries}\label{sec:prelim}\vspace{-2mm}
\xhdr{Hypergraph} A hypergraph $\mathcal{H}$ is a tuple $(V, \mathcal{E})$, where $V$ is a finite set of nodes, and $\mathcal{E}=\{E_1, E_2, ..., E_m\}$ is a set of sets with $E_i\subseteq V$ for all $1\leq i\leq m$. For the purpose of reconstruction, we assume the hyperedges are distinct, \textit{i.e.} $\forall 1\leq i, j\leq m,\;E_i\neq E_j$.

\vspace{-1mm}
\xhdr{Projected Graph} $\mathcal{H}$'s \textit{projected graph} (\textit{projection, \rv{Gaiman Graph \cite{yu1979algorithm}}}), $G$, is a graph with the same node set $V$, and (undirected) edge set $\mathcal{E}'$, \textit{i.e.} $G=(V, \mathcal{E}')$, where two nodes are joined by an edge in $\mathcal{E'}$ if and only if they belong to a common hyperedge in $\mathcal{E}$.  That is, \vspace{-1mm}
$$\mathcal{E}'=\{(v_i, v_j)|v_i, v_j \in E, E\in \mathcal{E}\}$$

\vspace{-2mm}
\xhdr{Maximal Cliques.} A \textit{clique} $C$ is a fully connected subgraph. We slightly abuse $C$ to also denote the set of nodes in the clique. A \textit{maximal clique} is a clique that cannot become larger by including more nodes. The \textit{maximal clique algorithm} returns all maximal cliques $\mathcal{M}$ in a graph \cite{bron1973algorithm}. The time complexity is linear to $|\mathcal{M}|$ \cite{tomita2006worst}. A \textit{maximum clique} is the largest maximal clique in a graph.  


\vspace{-0.5mm}
\section{Task Definition}\label{sec:def}
We define supervised hypergraph reconstruction as the following:
\begin{itemize}[leftmargin=3mm]
    \item \textbf{Input:} projected graph $G_1$, hypergraph $\mathcal{H}_0$; 
    \item \textbf{(Expected) output:} hypergraph $\mathcal{H}_1$;
    \item $\mathcal{H}_0$ and $\mathcal{H}_1$ belong to the same application domain, but their node indices are not aligned, \textit{i.e.} the learning is inductive.
    \item \textbf{Evaluaton:} following \cite{young2021hypergraph} we use the Jaccard Score as the main metric for evaluating reconstruction accuracy:
    \vspace{-2mm}$$\textbf{Jaccard Score}=\frac{|\mathcal{E}_1\cap\mathcal{R}_1|}{|\mathcal{E}_1\cup\mathcal{R}_1|}$$ \vspace{-1.5mm}
    $\mathcal{E}_1$ is the true hyperedges; $\mathcal{R}_1$ is the reconstructed hyperedges.
\end{itemize}

\vspace{-1mm}
\section{Topological Analysis}\label{sec:theory}
We start with a topological analysis to understand our task from a theoretical perspective. In particular, we seek to answer two questions: \textbf{Q1.} what topological properties make hypergraphs easy to reconstruct (so that the projection does not eliminate too much high-order  information)? \textbf{Q2.} in the extreme case, how much information is lost due to the projection?   By characterizing the notion of ``ease'' and ``difficulty'' in reconstruction, our analysis will help data contributors and researchers make more informed choice in dealing with high-order interactions.  


In principle, any clique in a projection qualifies as a candidate for a true hyperedge. Therefore, towards perfect reconstruction, we should consider $\mathcal{U}$, the universe of all cliques\hide{\footnote{Known as the \textit{clique complex}, an abstract simplicial complex in algebraic topology}} in $G$, including single nodes. To enumerate $\mathcal{U}$, a helpful view is the union of \rv{the power set} \footnote{\rv{Given a set $S$, its power set $\mathcal{P}(S)$ is defined as the set of all subsets of $S$. For example, $\{A, B, C\}$'s power set is $\{\{A, B, C\},\{A, B\},\{B, C\},\{A, C\},\{A\},\{B\},\{C\},\emptyset\}$.}}of each maximal clique of $G$. Mathematically,\vspace{-1mm}
$$\mathcal{U}=\bigcup_{C\in \mathcal{M}}\mathcal{P}(C)\setminus \emptyset$$
\vspace{-0.2mm}In that sense, maximal clique algorithm is a critical first step for hypergraph reconstruction, as was applied by \cite{young2021hypergraph} to initialize the state of its MCMC solver. \rv{In the extreme case, one can intuitively perceive} that if the hyperedges of $\mathcal{H}$ are mostly disjoint with each other, the maximal cliques in $\mathcal{H}$'s projection must be highly aligned with hyperedges of $\mathcal{H}$. It is impossible to find all hyperedges without considering all maximal cliques. \rv{Therefore, the accuracy of reconstruction by maximal clique algorithm is a good quantifier for measuring how ``difficult'' it is to reconstruct a hypergraph.}


{\parindent0pt
\begin{defn}\label{def:sperner}
A hypergraph $\mathcal{H}=(V, \mathcal{E})$ is \textit{Sperner} if for every hyperedge $E\in \mathcal{E}$ there does not exist a hyperedge $E'\in \mathcal{E}$ $s.t.$ $E\subset E'$.
\end{defn}
}

\vspace{-1mm}
{\parindent0pt
\begin{defn}\label{def:conformal}
A hypergraph $\mathcal{H}=(V, \mathcal{E})$ is \rv{\textit{conformal}} \footnote{\rv{Note that in database theory the important notions of $\alpha$-acyclicity \cite{beeri1983desirability} and GYO reducibility  \cite{yu1979algorithm}  extend conformity by further requiring the projected graph to be chordal. The interested readers are referred to the references for more details.}} if all the maximal cliques in its projection $G$ are its hyperedges, \textit{i.e.} $\mathcal{M}\subseteq\mathcal{E}$.
\end{defn}
}


\vspace{-1.5mm}
{\parindent0pt
\begin{theorem}\label{theorem:sperner_conformal}
The maximal cliques of $G$ are precisely all hyperedges of $\mathcal{H}$, \textit{i.e.} $\mathcal{M} = \mathcal{E}$, if and only if $\mathcal{H}$ is both Sperner and conformal.
\end{theorem}
}

\begin{figure}[]
    \centering
    \includegraphics[scale=0.57]{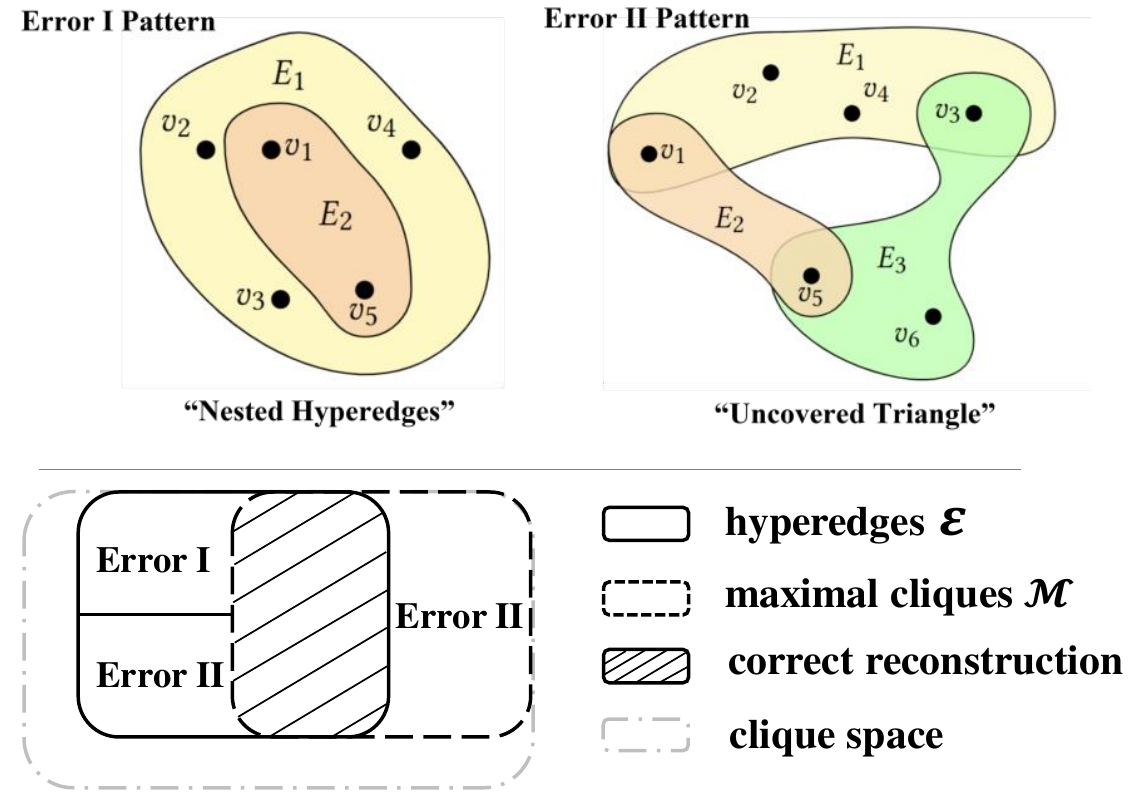}
    \vspace{-2.5mm}
    \caption{\small The upper half shows patterns of the two types of errors made by maximal cliques algorithm. \rv{In Error I pattern, $E_2$ will never be found (false negative); in Error II pattern, corners of the ``triangle'' are ``uncovered'', so the max clique $\{v_1, v_3, v_5\}$ is not a hyperedge (false positive), meanwhile $\{v_1, v_5\}$ is missed (false negative).} The bottom half shows the errors' relationship with $\mathcal{E}$ and $\mathcal{M}$. 
    If a hyperedge can't be reconstructed due to both errors, we count it as Error I.} 
    \label{fig:error_diagram}\vspace{-2mm}
\end{figure}

\vspace{-1mm}
\noindent Theorem \ref{theorem:sperner_conformal} gives the two strict conditions that $\mathcal{H}$ must satisfy in order to be ``easy'' to reconstruct. The Sperner's definition is self-explanatory --- simply summarized as "no pair of nested hyperedges"\hide{, illustrated by Fig.\ref{fig:error_diagram}, Error I Pattern}. In contrast, the conformal property is less clearer. The following theorem interprets the conformal property by its equivalence of "uncovered triangle" property.
\vspace{-0.5mm}
{\parindent0pt
\rv{\begin{theorem}\label{theorem:triangle}\vspace{-1mm}
A hypergraph $\mathcal{H}=(V, \mathcal{E})$ is conformal \textbf{iff} \hide{the pairwise intersections of every three hyperedges \rv{are} contained in a hyperedge, \textit{i.e.}:} for every three hyperedges there always exists some hyperedge $E$ such that all pairwise intersections of the three hyperedges are subsets of $E$, \textit{i.e.}:
\vspace{-1.5mm}
\begin{align*}
    \forall E_i, E_j, &E_q \in \mathcal{E},\\
    \; &\exists\;E\in \mathcal{E},\;s.t.\;(E_i \cap E_j) \cup (E_j \cap E_q) \cup (E_q \cap E_i) \subseteq E
\end{align*}
\end{theorem}}}
\vspace{-1mm}
\noindent The intuition behind Theorem \ref{theorem:triangle} is that a conformal hypergraph cannot have three hyperedges form a ``triangle'' whose three corners are ``uncovered''. \rv{The theorem gives a nontrivial interpretation of 
Def. \ref{def:conformal} by eliminating the notion of cliques or max cliques. It shows how to check a hypergraph’s conformity purely from its hyperedge patterns with no need to further compute any max cliques, which is a more intuitive way for us to visually understand conformity.
} \hide{The Error II Pattern in Fig.\ref{fig:error_diagram} illustrates this idea.}

\rv{
Based on Def. \ref{def:sperner} and \ref{def:conformal}, we further define the two types of errors that the maximal cliques algorithm makes:
{\parindent0pt
\begin{defn}\label{def:errors}\vspace{-1mm}
    There exist two types of Error If we treat all maximal cliques as true hyperedges: an error is defined as \textbf{ Error I (Error II)} if it occurs because of the projected graph's violation of Def. \ref{def:sperner} (Def. \ref{def:conformal}).\vspace{-1mm}
\end{defn}}

Fig.\ref{fig:error_diagram} illustrates patterns of the two errors  and their relationship with other important concepts in our task. Note that here Error I and II are different from the well-known Type I (false positive) and Type II (false negative) Error In statistics. In fact, Error I’s are hyperedges that nest inside some other hyperedges, so they are indeed false negatives; Error II’s can be either false positives or negatives. For example, in Fig.\ref{fig:error_diagram} “Error II Pattern”  : $\{v_1, v_3, v_5\}$ is a false positive, $\{v_1, v_5\}$ is a false negative.}

\begin{table}[]
    \centering
\resizebox{\columnwidth}{!}{%
\begin{tabular}{lcccccc}
\hline
\textbf{Dataset} & $|\mathcal{E}|$ & $|\mathcal{E}'|$ & $|\mathcal{M}|$ & \textbf{Error I} & \textbf{Error II} & \textbf{} \\ \hline
DBLP \cite{benson2018simplicial}       & 197,067 & 194,598 & 166,571 & 2.02\% & 18.9\% &  \\
Enron  \cite{benson2018simplicial}      & 756     & 300     & 362     & 42.5\% & 53.3\% &  \\
Foursquare \cite{young2021hypergraph} & 1,019   & 874     & 8,135   & 1.74\% & 88.6\% &  \\
Hosts-Virus \cite{young2021hypergraph}      & 218     & 126     & 361     & 19.5\% & 58.1\% &  \\
H. School \cite{benson2018simplicial} & 3,909   & 2864    & 3,279   & 14.9\% & 82.7\% &  \\ \hline
\end{tabular}
}
\caption{\small $\mathcal{E}$ is the set of hyperedges; $\mathcal{E}'$ is the set of hyperedges not nested in any other hyperedges; $\mathcal{M}$ is the set of maximal cliques in $G$. \textbf{Error I, II} result from the violation of conformal and Sperner properties, respectively. Error I $=\frac{|\mathcal{E}\backslash\mathcal{E}'|}{|\mathcal{E}\cup\mathcal{M}|}$, Error II $=\frac{|\mathcal{M}\backslash\mathcal{E}'|+ |\mathcal{E}'\backslash\mathcal{M}|}{|\mathcal{E}\cup\mathcal{M}|}$.}
\label{tab:errors}
\vspace{-7mm}
\end{table}


Because both patterns in Fig.\ref{fig:error_diagram} can be easily realized, real-world hypergraphs can easily deviate from either / both of the two properties. The significance of deviation is closely associated with the error rate of maximal cliques algorithm in reconstruction. A hypergraph that violates the Sperner property badly can have many pairs of ``nested hyperedges''. Common examples include hypergraphs of email correspondence and social interactions. See Table \ref{tab:errors}, Error I column. In the extreme case, a hypergraph contains one large hyperedge plus many nested hyperedges as proper subsets. Therefore, the worst-case accuracy of maximal cliques algorithm on a non-Sperner but conformal hypergraph is $1/(2^n-1)$.

\noindent That said, one may argue there are still many real-world hypergraphs that are Sperner or almost Sperner. It turns out that the worst case of violating the conformal property can also be disastrous:

\vspace{-1mm}
{\parindent0pt\begin{theorem}\label{theorem:count}
Let $\mathcal{H}=(V,\mathcal{E})$ be Sperner with $m=|\mathcal{E}|$, and $p^{(\mathcal{H})}$ the accuracy of maximal clique algorithm for reconstructing $\mathcal{H}$, then\vspace{-1mm}
$$\min_{\mathcal{H}} p^{(\mathcal{H})}\leq 2^{-{m-1 \choose [m/2]-1}}\ll 2^{-m}$$
\end{theorem}}

\hide{
\vspace{-4mm}
{\parindent0pt\begin{sproof}
Because $\mathcal{H}$ is Sperner, each hyperedge must be a maximal clique. Define $\lambda(m)=1/p^{(\mathcal{H})}$, then we are equivalently to prove $\lambda(m)\geq2^{{m-1 \choose [m/2]-1}}$. Here the physical meaning of $\lambda(m)$ is the number of maximal cliques that can be generated by $m$ hyperedges (that are inclusion-wise maximal).

We show that $\lambda(m)$ is equal to the number of maximal families of mutually intersecting subsets of $X=\{1, 2, ..., m\}$. A maximal family of mutually intersecting subsets of $X$, $\mathcal{F}$, is a collection of $X$'s subsets in which (1) the pairwise intersection of any two subsets (not necessarily distinct) is non-empty, and (2) no other subset of $X$ can be added. We can prove that each maximal clique corresponds to a unique $\mathcal{F}$. Counting the exact number of $\mathcal{F}$'s w.r.t. $m$ remains unsolved to date, but a lower bound $2^{{m-1 \choose [m/2]-1}}$ can be obtained, which completes our proof. See \ref{subsec:count} for details.
\end{sproof}}
}
\vspace{-2mm}
While the worst case rarely arises in practice, real-world hypergraphs often create many maximal cliques that are not hyperedges due to the easy construction of Error II patterns. 

\vspace{-1mm}
\xhdr{Summary} With the topological analysis, we obtain a clearer picture of the types of hypergraphs that are easy/hard to reconstruct. The easy construction of the two error patterns indicates that the reconstruction errors can extensively occur in real-world datasets. It is extremely hard to fix these errors by just looking at the projection, which necessitates the usage of a training hypergraph. How do we make the best of the training hypergraph to help with reconstruction? We elaborate this in the next section. 


\vspace{-1mm}
\section{Reconstruction Framework}\label{sec:framework}
\subsection{Overview}
\begin{figure}[]
    \centering\vspace{-2mm}
    \includegraphics[scale=0.42]{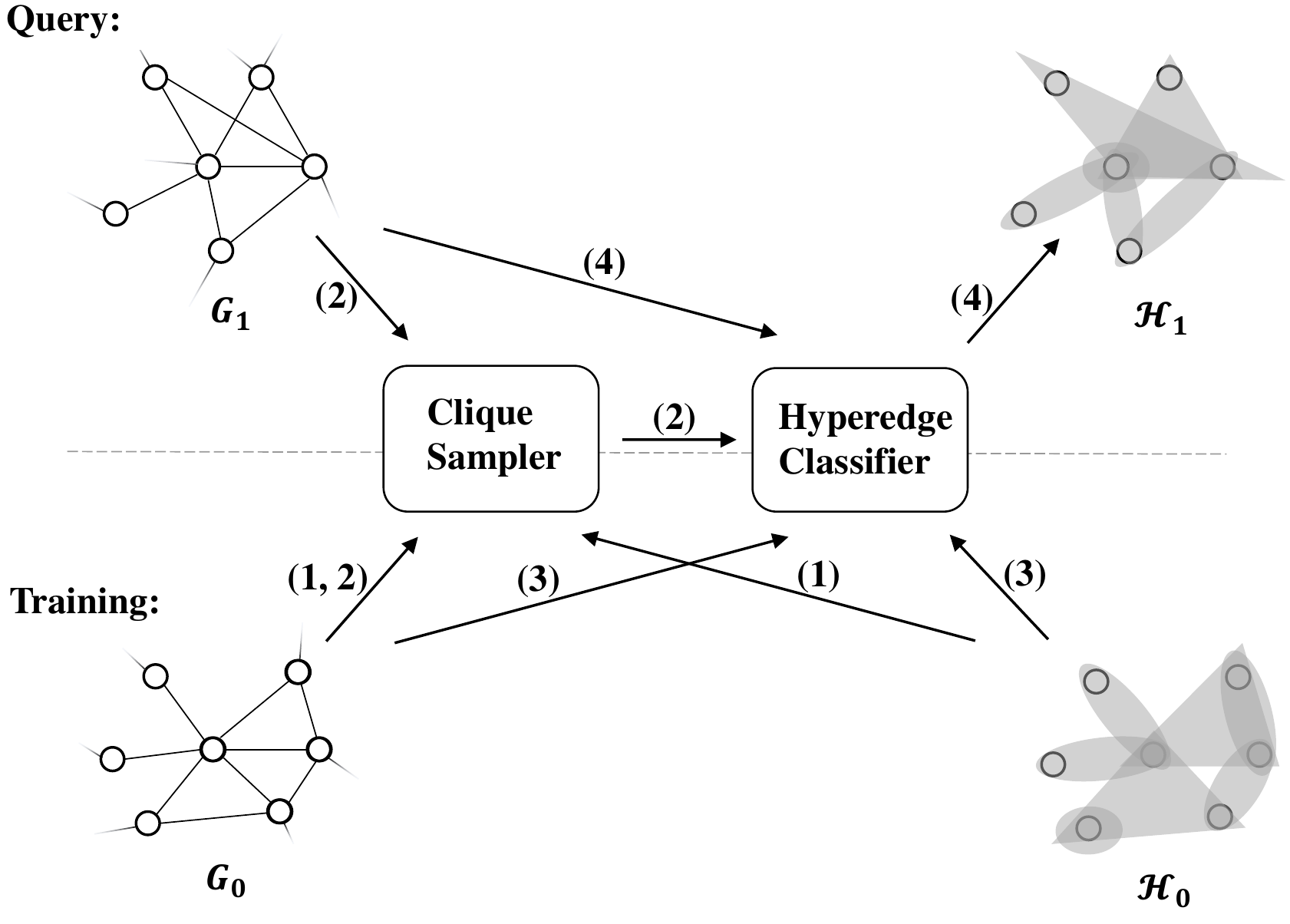}
    \vspace{-4mm}
    \caption{\footnotesize Our reconstruction framework takes 4 steps: (1) the clique sampler is optimized on $G_0$ and $\mathcal{H}_0$; (2) the clique sampler samples candidates from $G_0$ and $G_1$, and passes the result to the hyperedge classifier; (3) the hyperedge classifier extracts features of candidates from $G_0$ and trains on them; (4) the hyperedge classifier extracts features of candidates from $G_1$ and identify hyperedges.}\vspace{-4.5mm}
    \label{fig:framework}
\end{figure}
We have established that a hypergraph is hard to reconstruct purely based on its projection if it contains many hyperedge overlaps of the two patterns. In this section, we solve this by leveraging supervised signals. The high-level idea is that we use a \textit{clique sampler} to narrow the search space of hyperedges (clique space in Fig.\ref{fig:error_diagram}), and then use a \textit{hyperedge classifier} to identify hyperedges from the narrowed space. Both modules are optimized/trained on the training data.

Fig.\ref{fig:framework} gives a more detailed 4-step view, briefly explained in the caption. They will be elaborated in the following subsections: Sec. \ref{subsec:consistency} introduces an important observation that underpins the clique sampler; Sec. \ref{subsec:sampler} details the clique sampler and its optimization; Sec. \ref{subsec:classifier} expalins principles and designs for the hyperedge classifier.

\vspace{-1.5mm}
\subsection{$\rho(n, k)$-consistency}\label{subsec:consistency}
The $\rho(n, k)$-consistency describes the consistency that we observe in hyperedge distribution with regard to maximal cliques, among hypergraphs from the same application domain, \textit{e.g.} $\mathcal{H}_0$, $\mathcal{H}_1$. It is an important property that we leverage in supervised reconstruction.

Given a hypergraph $\mathcal{H}=(V, \mathcal{E})$, its projection $G$, maximal cliques $\mathcal{M}$, we use $\rho(n, k)$ to denote the expectation that we find a unique hyperedge by randomly sampling a size-$k$ subset from a random size-$n$ maximal clique. A $(n,k)$ is called \textit{valid} if $1\leq k\leq n\leq N$, with $N$ the size of the \textit{maximum} clique. $\rho(n, k)$ can be estimated empirically via the unbiased estimator $\hat{\rho}(n, k)$ defined as:\vspace{-1.5mm}
$$\hat{\rho}(n, k)=\frac{|\mathcal{E}_{n,k}|}{|\mathcal{Q}_{n,k}|}$$
\vspace{-3.5mm}where
\vspace{-2.5mm}\begin{align*}
    \mathcal{E}_{n,k} &= \{S\in \mathcal{E}|S\subseteq C,|S|=k, C\in \mathcal{M},|C|=n\} \\
    Q_{n,k} &=\{(S,C)|S\subseteq C, |S|=k, C\in \mathcal{M},|C|=n\}\vspace{-5mm}
\end{align*}
\noindent$\mathcal{E}_{n,k}$ denotes the set of size-$k$ hyperedges in size-$n$ maximal cliques. $Q_{n,k}$ denotes all possible ways to sample a size-$n$ maximal clique and then a size-$k$ subset (\textit{i.e.} a $k$-clique) from the maximal clique. $|Q_{n,k}|$ can be further simplified as $|\{C|C\in \mathcal{M},|C|=n\}|{n \choose k}$.

\begin{figure}[]
     \begin{subfigure}[b]{0.5\textwidth}
    \includegraphics[scale=0.29]{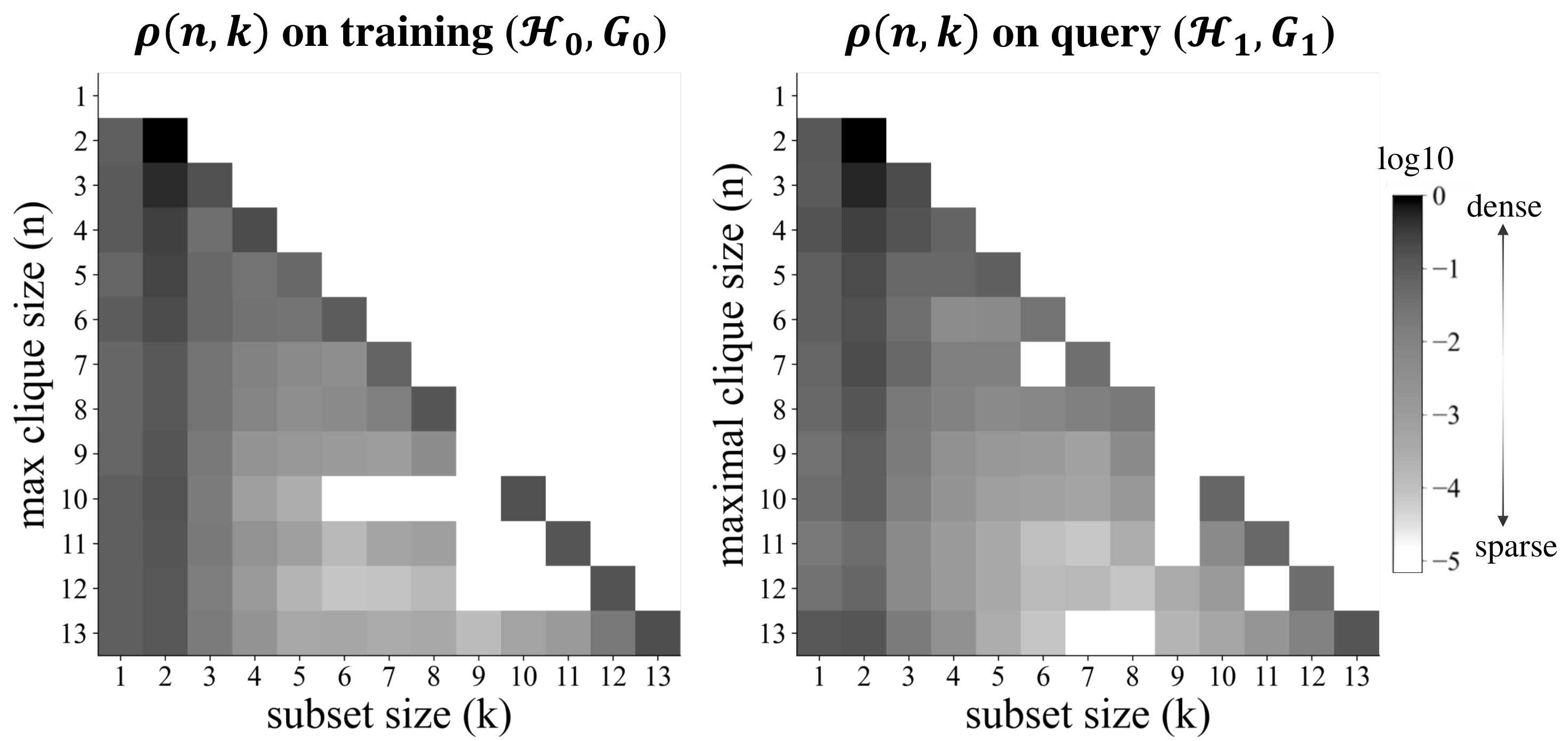}\vspace{-2mm}
    \caption{}\label{fig:consistency}
    \end{subfigure}
    \begin{subfigure}[b]{0.5\textwidth}
    \includegraphics[scale=0.31]{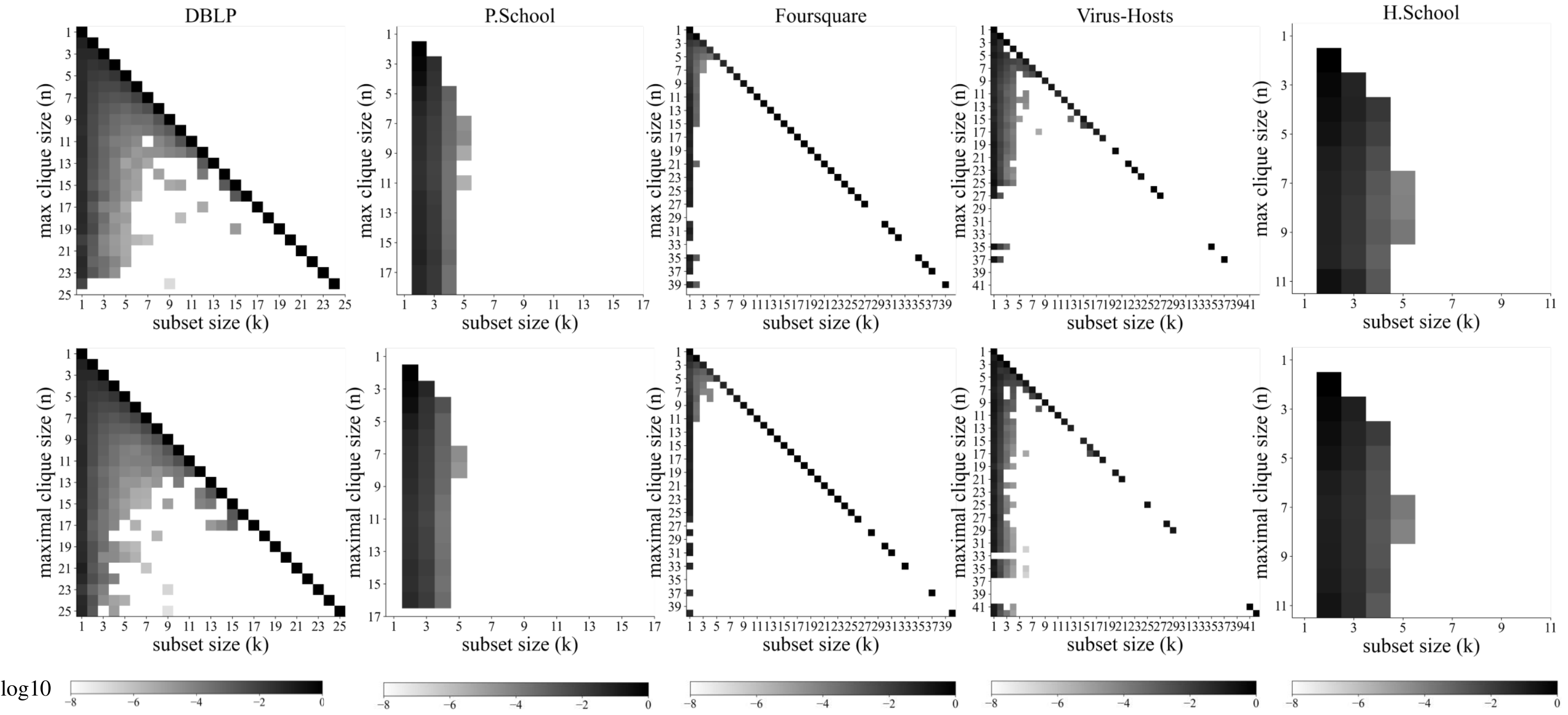}\vspace{-2mm}
    \caption{}\label{fig:more_rho}
    \end{subfigure}\vspace{-3mm}
    \caption{\small{(a) $\rho(n, k)$-consistency on dataset Enron \cite{benson2018simplicial}, where each node is an email address, each hyperedge is an email. $\mathcal{H}_0$ and $\mathcal{H}_1$ are obtained by splitting all emails by a middle timestamp. \rv{(b) $\rho(n, k)$-consistency on more datasets. Notice that heatmaps in the same columns are similar, while those in the same row are much different.}}}\vspace{-6mm}
\end{figure}

Our key observation is that if two hypergraphs \textit{e.g.} $\mathcal{H}_0$, $\mathcal{H}_1$, are generated from the same source, their hyperedge distributions admit similar distributions of $\rho(n, k)$, which we call \textit{$\rho(n, k)$-consistency}. Fig.\ref{fig:consistency} uses heatmaps to visualizes $\rho(n, k)$-consistency on a famous communication dataset, Enron \cite{benson2018simplicial}, where $\mathcal{H}_0$ and $\mathcal{H}_1$ are split based on a median timestamp of the emails (hyperedges). $n$ is plot on the $y$-axis, $k$ on the $x$-axis. \rv{Fig.\ref{fig:more_rho} further plots $\rho(n, k)$'s heatmaps for five other datasets in a similar manner. From both figures, we observe that the distributions of $\rho(n,k)$ exhibit good consistency between training and query of the same dataset; in contrast, the distributions of $\rho(n,k)$ across datasets are much different. This visual observation can be further confirmed by quantified measures (see Fig.\ref{fig:distance} in Appendix).} 

Also, notice that the second column and the diagonal are darkest, \rv{implying that the ${n \choose k}$ term in $Q_{n,k}$ (the denominator) cannot dominate the distribution. More concretely, ${n \choose k}$ reaches minimum when $k=n$ or $1$ and grows exponentially as $k$ approaches $0.5n$, and this is true regardless of the data; here the quotient peaking at $k=2$ means that the term $|\mathcal{E}_{n,k}|$ that reflects data is playing a numerically meaningful role.} By examining many datasets (see Fig.\ref{fig:more_rho}), we find that the $\rho(n, k)$ distribution can vary a lot, but there is always great consistency between training hypergraphs and query hypergraphs from the same application domain.

\yb{\noindent\textbf{Complexity.} The main complexity of computing $\rho(n, k)$ involves two parts: \textit{(\textbf{a})} computing $\mathcal{M}$; \textit{(\textbf{b})} computing $\mathcal{E}_{n,k}$ for all valid $(n,k)$. 

\textit{(\textbf{a})}'s complexity is $O(|\mathcal{M}|)$ as mentioned in Sec. \ref{sec:prelim}. Though in worst case $|\mathcal{M}|$ can be exponential to $|V|$, in practice we often observe $|\mathcal{M}|$ on the same magnitude order as $|\mathcal{E}|$ (see Table \ref{tab:errors}), which is an interesting phenomenon. 

\textit{(\textbf{b})} requires matching size-$n$ maximal cliques with size-$k$ hyperedges for each valid $(n,k)$ pair. The key is that in real-world data, the average number of hyperedges incident to a node is usually a constant independent from the growth of $|V|$ or $|\mathcal{M}|$ (see also \textbf{$\bar{d}(V)$} in Table. \ref{tab:dataset}), known as the ``sparsity of hypergraphs''\cite{kook2020evolution}. This property greatly reduces the (average size of) search space for all size-$k$ hyperedges in a size-$n$ maximal clique from $|\mathcal{E}|$ to $n\bar{d}(V)$. As we see both $n$ and $\bar{d}(V)$ are typically under $50$ in practice and neither grows with $|\mathcal{M}|$, \textit{\textbf{b}}'s complexity can still be viewed as $O(|\mathcal{M}|)$. Therefore, the total complexity for computing $\rho(n,k)$ is $O(|\mathcal{M}|)$.}
We extend this result with more empirical evidence in Sec. \ref{subsec:exp_quality}

\hide{
\vspace{1mm}
\noindent$\rho(n, k)$-consistency allows us to design a clique sampler to be applied over $G_1$, using knowledge of $\rho(n, k)$ in $\mathcal{H}_0$ and $G_0$.}


\vspace{-1mm}
\subsection{Clique Sampler}\label{subsec:sampler}
Given a query graph, usually we cannot afford to take all its cliques as candidates for hyperedges\rv{: although we just mentioned that the number of maximal cliques $|\mathcal{M}|$ is often manageable, the number of cliques $|\mathcal{U}|$ can far exceed handling capacity as one size-$n$ maximal clique produces $2^n-1$ cliques.} Therefore, we create a clique sampler. Assume we have a limited sampling budget $\beta$, our goal is to collect as many hyperedges as possible by sampling $\beta$ cliques from $\mathcal{U}$. Any hyperedge missed by our clique sampler loses the chance to be identified by the hyperedge classifier in the future, so this step is crucial. \rv{Later we show in Table \ref{tab:beta} that good performance can be achieved by $\beta$'s that are orders of magnitude smaller than |$\mathcal{U}$|.}

As mentioned, by looking at just the query $G_1$ we cannot locate hyperedges in the enormous search space of all cliques in $G_1$. Fortunately, we can obtain hints from $G_0$ and $\mathcal{H}_0$. The idea is that we use $G_0$ and $\mathcal{H}_0$ to optimize a clique sampler that verifiably collects a good number of hyperedges. The optimization process, as marked in Fig.\ref{fig:framework} step 1, can be viewed as a procedure that learns knowledge about where to sample. Then in $G_1$, we use the optimized clique sampler to sample cliques, shown in Fig.\ref{fig:framework} step 2. We ask that the sampler takes the following form: 
\begin{itemize}[leftmargin=5mm]
    \item[\textbf{($\star$)}] For each valid $(n,k)$, we sample a total of $r_{n,k}|Q_{n,k}|$ size-$k$ subsets (\textit{i.e.} $k$-cliques) from size-$n$ maximal cliques in the query graph, subject to the sampling budget: $\sum_{n,k}{r_{n,k}|Q_{n,k}|}=\beta$
\end{itemize}
$r_{n,k}\in[0,1]$ is the sampling ratio of the $(n,k)$ cell, $|Q_{n,k}|$ is the size of the $(n,k)$ cell's sample space in $G_0$. To instantiate a sampler, a $r_{n,k}$ should be specified for every valid $(n,k)$ cell. How to determine $r_{n,k}$'s? We optimize $r_{n,k}$'s towards collecting the most training hyperedges from $G_0$, with the objective:\vspace{-1mm}
\begin{align*}
    \{r_{n,k}\}&=\argmax_{\{r_{n,k}\}}\;\eoc[ \bigcup_{(n,k)}r_{n,k}\odot\mathcal{E}_{n,k}]
    \vspace{-3.5mm}
\end{align*}
$\odot$ is a \textit{set sampling} operator that returns a uniformly downsampled subset of $\mathcal{E}_{n,k}$ at downsampling rate $r_{n,k}$. $\odot$ is essentially a generator for \textit{random finite set} \cite{mullane2011random}. See \ref{subsec:proof_sampler} for more discussion. $\eoc[\cdot]$ returns the expected cardinality. Given the maximization objective and $\rho(n, k)$-consistency, the optimized sampler should also collect a good number of hyperedges when generalized to $G_1$. Sec.\ref{subsec:exp_abl} validates this claim empirically.

\noindent\textbf{Optimization.} To collect more hyperedges from $G_0$, a heuristic is to allocate all budget to the darkest cells of the training data's heamap (Fig.\ref{fig:consistency}-left), where hyperedges most densely populate. However, a caveat is that the set of hyperedges $\mathcal{E}_{n,k}$ in each $(n,k)$ cell are not disjoint if the cells are in the same column. In other words, a size-$k$ clique can appear in multiple maximal cliques of different sizes. Therefore, taking the darkest cells may not yield best result. In fact, optimizing the objective above involves maximizing a monotone submodular function under budget constraint, which is NP-hard.

In light of this, we design the greedy Alg. \ref{algo:sampler} to approximate the optimal solution with worst-case guarantee. It takes four inputs: sampling budget $\beta$, size of the \textit{maximum} clique $N$, $\mathcal{E}_{n,k}$ and $\mathcal{Q}_{n,k}$ for all $1\leq k\leq n \leq N$. Lines \ref{line:init_begin}-\ref{line:init_end} initialize state variables. Lines \ref{line:greedy_begin}-\ref{line:greedy_end} run greedy selection iteratively.  

\vspace{-1mm}
\begin{algorithm}
\caption{Optimize Clique Sampler}\label{algo:sampler}
\begin{algorithmic}[1]
\Require{$\beta$; $N$; $\mathcal{E}_{n,k}$, $\mathcal{Q}_{n,k}$ for all $1\leq k\leq N, k\leq n \leq N$}
\For{$k=1$ to $N$}\label{line:init_begin}\Comment{\small traverse $k$ to initialize state variables}
    \State{$\Gamma_k\gets \emptyset$}\Comment{\small union of $\mathcal{E}_{n,k}$'s picked from column $k$}\label{line:init_gamma}
    \State{$\omega_k\gets \{k, k+1, ..., N\}$}\Comment{\small available column-$k$ cells}\label{line:init_omega}
    \State{$r_{i,k}\gets 0$ for $i\in\omega_k$}\label{line:init_r} \Comment{\small sampling ratios for column-$k$ cells}
    \State{$\Delta_k,n_k\gets\;$\update($k$, $\omega_k$, $\Gamma_k$, $\mathcal{E}_{\cdot,k}$, $\mathcal{Q}_{\cdot,k}$)}\label{line:init_delta_n}
\EndFor\label{line:init_end}
\While{$\beta>0$}\label{line:greedy_begin}\Comment{\small the greedy selection starts}
    \State{$k\gets \text{argmax}_{i}{\;\Delta_i}$}\label{line:greedy_pick}\Comment{selects the next best $k$}
    \State{$r_{n_k,k}\gets \min\{1, \frac{\beta}{|\mathcal{Q}_{n_k, k}|}\}$}\label{line:cell}\Comment{\small sets cell's sampling ratio}
    \State{$\Gamma_k\gets \Gamma_k \cup \mathcal{E}_{n_k, k}$}\Comment{\small updates state variables}
    \State{$\omega_k\gets\omega_k\backslash\{n_k\}$}
    \State{$\beta\gets\beta-|\mathcal{Q}_{n_k, k}|$}
    \State{$\Delta_k,n_k\gets\;$\update($k$, $\omega_k$, $\Gamma_k$, $\mathcal{E}_{\cdot,k}$, $\mathcal{Q}_{\cdot,k}$)}
    \State{\textbf{if} $\text{max}_{k}\;\Delta_k=0$ \textbf{then} break}\Comment{\small breaks if all cells sampled}
\EndWhile\label{line:greedy_end}
\State \Return{$r_{n,k}$ for all $1\leq k\leq N, k\leq n \leq N$} 
\end{algorithmic}
\end{algorithm}
\setcounter{algorithm}{0}
\makeatletter
\renewcommand*{\ALG@name}{Subroutine}
\makeatother
\vspace{-5mm}
\begin{algorithm}
\caption{\update}
\begin{algorithmic}[1]
\Require{$k$; $\omega_k$; $\Gamma_k$; $\mathcal{E}_{\cdot,k}$; $\mathcal{Q}_{\cdot,k}$}
\If{$\omega_k\neq \emptyset$}
    \State{$\Delta'\gets \max_{n\in\omega_k}
    \frac{|\Gamma_k\cup\mathcal{E}_{n,k}|-|\Gamma_k|}{|\mathcal{Q}_{n,k}|}$}
    \State{$n'\gets \text{argmax}_{n\in\omega_k}
    \frac{|\Gamma_k\cup\mathcal{E}_{n,k}|-|\Gamma_k|}{|\mathcal{Q}_{n,k}|}$}
\Else
    \State{$\Delta'\gets 0$; $n'\gets 0$;}
\EndIf
\State{\Return{$\Delta'$, $n'$}}
\end{algorithmic}
\end{algorithm}

The initialization is done column-wise. In each column $k$, $\Gamma_k$ stores the union of all $\mathcal{E}_{n,k}$ that have been selected from column $k$ so far; $\omega_k$ stores the row indices (\textit{i.e.} $n$'s) of all available cells in column $k$, \textit{i.e.} cells that have not been selected; $r_{i,k}$ is the sampling ratio of each valid cell;  line \ref{line:init_delta_n} calls the subroutine \update to compute $\delta_k$, which is the best sampling efficiency among all the available cells, and $n_k$, which is the row index of that most efficient cell.

Lines \ref{line:greedy_begin}-\ref{line:greedy_end} execute the greedy selection. In each iteration, we greedily take the next most efficient cell among (the best of) all the columns, store the decision in the corresponding $r$, and update ($\Gamma$, $\omega$, $\delta_k$, $n_k$) with $k$ the column index of the selected cell in the current iteration. Notice that only column $k$ needs to be updated because $\mathcal{E}_{n,k}$'s with different $k$'s are independent. Finally, the greedy stops when either having exceeded budget or having traversed all cells. 

For a more intuitive view, please refer to Sec.\ref{subsec:exp_abl} where we do ablations on this algorithm and visualize the iterations. As a side bonus, we will also see in Sec. \ref{subsec:exp_quality} that the greedy actually induces more diverse hyperedge sizes in reconstruction. 

\vspace{3.5mm}
{\parindent0pt
\begin{theorem}\label{theorem:sampler}
Let $q$ be the expected number of hyperedges in $\mathcal{H}_0$ drawn by the clique sampler optimized by Alg. \ref{algo:sampler}; let $q^*$ be the expected number of hyperedges in $\mathcal{H}_0$ drawn by the best-possible clique sampler, with the same $\beta$. Then,\vspace{-3mm} $$q>(1-\frac{1}{e})q^*\approx 0.63q^*$$
\end{theorem}}
\vspace{-1.2mm}\noindent Theorem \ref{theorem:sampler} gives an important quality guarantee on Alg. \ref{algo:sampler} by comparing its result with the best-possible clique sampler. The actual $\frac{q}{q^*}$ can often be much better than $0.63$. Besides, notice that Alg. \ref{algo:sampler} leaves at most one $(n,k)$ cell partially sampled. \textbf{\textit{Is that a good design?}} In fact, it can be proved that there is always one best clique sampler that leaves at most one $(n,k)$ cell partially sampled. Otherwise, we can always find two partially sampled cells and relocate all our budget from one to the other to achieve a higher $q$.

\noindent\rv{\textbf{Relating to Error I \& II.} The effectiveness of our clique sampler can also be understood from the perspective of reducing Error I and II. Taking Fig.\ref{fig:consistency} as an example: by learning which non-diagonal cells to sample, the clique sampler essentially reduces Error I as well as the false negative part of Error II; by learning which diagonal cells to sample, it further reduces the false positive part of Error II.}

\noindent\rv{\textbf{Relating to Standard Submodular Optimization.}Our clique sampler takes a similar greedy form as the standard algorithm for submodular optimization. However, our clique sampler actually solve a more challenging problem, whose optimality guarantee is also harder to prove. Please see \ref{subsec:standard_submodular} for more details.} 

\noindent\textbf{Precision-Recall Tradeoff.} For each dataset, $\beta$ should be specified by humans. What is the best $\beta$? Clearly a larger $\beta$ yields a larger $q$, and thus a higher recall $\frac{q}{|\mathcal{E}|}$ with our samples. On the other hand, a larger $\beta$ also means a lower precision $\frac{q}{\beta}$, as sparser regions get sampled. $\frac{q}{\beta}$ being overly low harms sample quality and may jeopardize the training later. Such tradeoff calls for more calibration of $\beta$. We empirically found it often good to search $\beta$ in a range that makes $\frac{q}{|\mathcal{E}|}\in[0.6,0.95]$. More tuning details are in \ref{sec:more_exp}.


\noindent\textbf{Complexity.} The bottleneck of Alg. \ref{algo:sampler} is \update. In each iteration after a $k$ is picked, \update recomputes $(|\Gamma_k\cup\mathcal{E}_{n,k}|-|\Gamma_k|)$ for all $n\in\omega_k$, which is $O(\frac{|\mathcal{E}|}{N})$. Empirically we found the number of iterations under the best $\beta$ always $O(N)$. $N$ is the size of the \textit{maximum} clique, and mostly falls in $[10,40]$ (see Fig.\ref{fig:more_rho}). Therefore, on expectation we would traverse $O(N)O(\frac{|\mathcal{E}|}{N})=O(|\mathcal{E}|)$ hyperedges if $|\mathcal{E}_{n,k}|$ distributes evenly among different $k$'s. In the worst case where $|\mathcal{E}_{n,k}|$'s are deadly skewed, this degenerates to $O(N|\mathcal{E}|)$,

The optimized sampler samples $\beta$ cliques from $G_0$, and a similar number of cliques from $G_1$. Later in Fig.\ref{fig:exp_sampler}, we empirically confirm that the two sampling sizes are similar indeed.

\vspace{-3mm}
\subsection{Hyperedge Classifier}\label{subsec:classifier}
 A \textit{hyperedge classifier} is a binary classification model that takes a target clique in the projection as input, and outputs a 0/1 label indicating whether the target clique is a hyperedge. We train the hyperedge classifier on $(\mathcal{H}_0, G_0)$ where we have ground-truth labels (step 3, Fig.\ref{fig:framework}), and then use it to identify hyperedges from $G_1$ (step 4, Fig.\ref{fig:framework}). To serve this purpose, a hyperedge classifier should contain two parts (1) a feature extractor that extracts expressive features for characterizing a target clique as a subgraph, and (2) a binary classification model that takes an extracted feature and outputs a 0/1 label. (2) is a standard task and so we use MLP with one layer of $h=100$ hidden neurons. (1) requires more careful design. Next, we first discuss (1)'s design principles, then provide two realizations that we found work well. \rv{Our hyperedge classifier addresses Error I and II together by fitting to the distribution of hyperedges with structural features of the clique candidates as input.}

\vspace{-3mm}
\subsubsection{\textbf{Design Principles}}
Central to the design of a good feature extractor is a question: what type of information should it capture about a target clique in the projection? In our task setup, there are two constraints usable information to ensure broadest usability of our approach: (1) we do not assume nodes to be attributed, or edges to have multiplicity; (2) we focus on the inductive case where node identities are not aligned across training and query (so positional embeddings like DeepWalk \cite{perozzi2014deepwalk} is unhelpful).

It thus becomes clearer that we are left with the structural features \textit{of} and \textit{around} the target clique. Put differently, the classifier needs to well characterize the connectivity properties of the local ego graph centered at the target clique. As we will see in experiment, using structural features in a supervised manner boosts reconstruction accuracy by an order of magnitude on difficult datasets. 

In that regard, any structural learning model that can well characterize the structural features can potentially be the classifier --- and there are plenty of them operating on individual nodes \cite{henderson2012rolx,li2020distance,xu2018powerful}. However, what we do not know yet is what structural features that characterize a clique (and its surroundings) are important, and how they can be understood by humans in the clique's context. Therefore, we provide the following two feature extractors that accomplish the task via interpretable features. It is worth mentioning though that they are not the only choices. 

\vspace{-2mm}
\subsubsection{\textbf{``Count''-based Feature Extractor}}
Many powerful graph structural learning models use different notions of ``count'' to characterize the local connectivity patterns. For example, GNNs typically use node degrees as initial features when node attributes are unavailable; the Weisfeiler-Lehman Test also updates a node's color based on the count of different colors in its neighborhood. 

In the context of a target clique as a subgraph structure in a projection, the notion of ``count'' can be especially rich in meaning: a target clique can be characterized by the count of its [own nodes / neighboring nodes / neighboring edges / attached maximal cliques] in many different ways. Therefore, we create a total of 8 types of generalized count-based features, detailed in \ref{subsec:more_count}. Despite being conceptually simple, these count features works surprisingly well and can be easily interpreted (see more in Supplement \ref{sec:more_exp}).

\vspace{-2mm}
\subsubsection{\textbf{Clique-motif-based Feature Extractor}}
\begin{figure}
    \centering\vspace{-1.5mm}
    \includegraphics[scale=0.54]{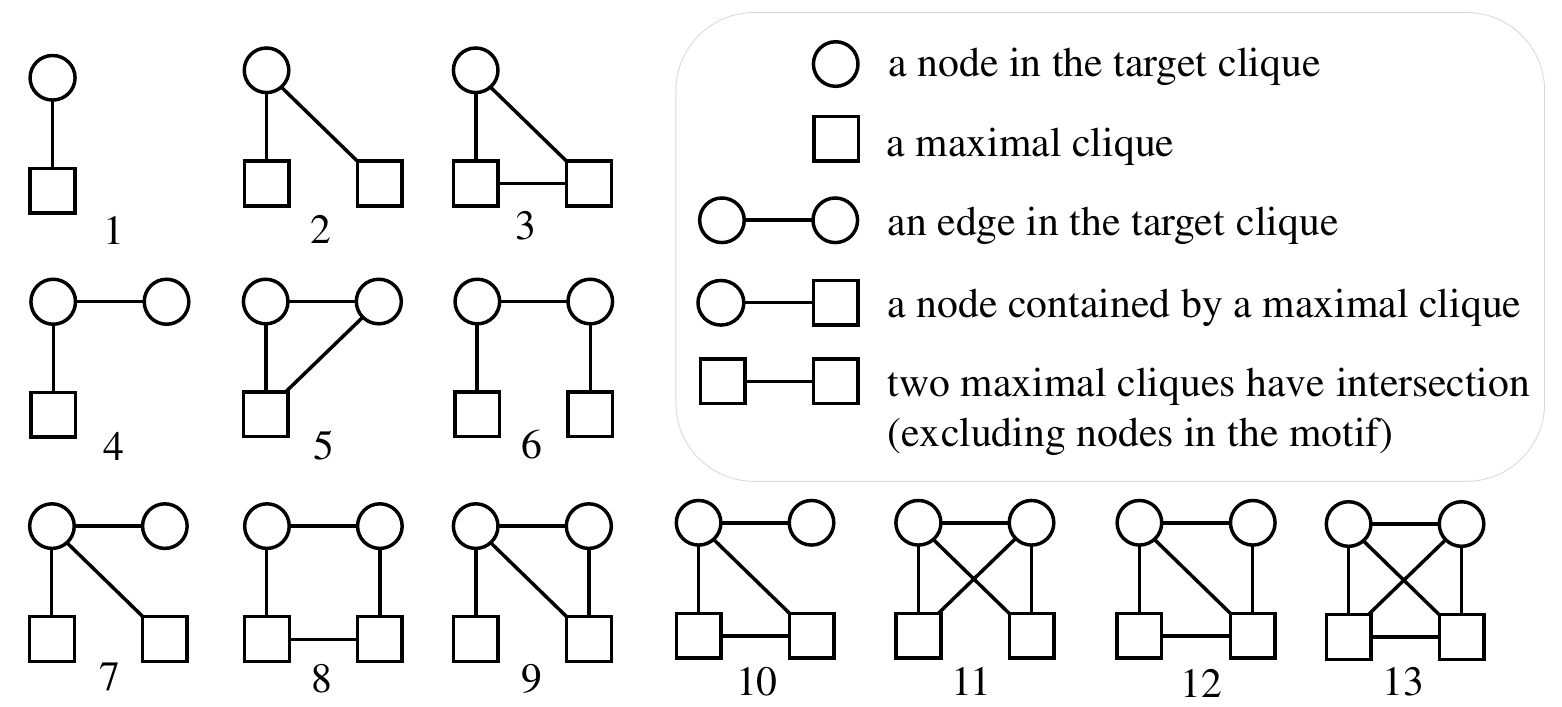}\vspace{-2.5mm}
    \caption{\small 13 ``clique motifs'' that encode rich connectivity patterns around a target clique $C$. Each clique motif is formed by 1 or 2 nodes in the target clique $+$ 1 or 2 maximal cliques that contain the nodes.}
    \label{fig:motifs}\vspace{-5mm}
\end{figure}
As a second attempt we use maximal cliques as an intermediate that bridges the projection and the hyperedges. The rationale behind is that the maximal cliques of a projection can essentially be viewed as a preliminary guess of the projection's high-order structures. Such guess not only encompasses all the projection's information, but also in itself constitutes a partially curated form of higher-order structures (which are indicative of the real higher-order structures, \textit{i.e.} hyperedges). Meanwhile, the synergy between maximal cliques and nodes in the target clique also exhibit rich connectivity patterns, which generalizes the notion of motif, as shown in Fig.\ref{fig:motifs}. 
    
\rv{Fig.\ref{fig:motifs} enumerates the all 13 connectivity patterns between 1 or 2 nodes of the target clique and 1 or 2 maximal cliques that contain the target clique's nodes.}  For simplicity we restrict that maximal cliques are at most one hop away from the target clique. We call the connectivity patterns \textbf{clique motifs}. \rv{Compared to count features, clique motifs are a more principled and systematic way to extract structural properties.} There are two types of components in a clique motif: node, maximal clique, and three types of relationships: node-node, node-maximal clique, maxmimal clique-maximal clique. Their concrete meanings are explained in the legend. We say that a clique motif is \textit{attached} to a target clique $C$ if the clique motif contains at least one node of $C$. We further use $\Phi_i^{\mathsmaller{(C)}}$ to denote the set of type-$i$ clique motifs attached to $C$, with $1\leq i\leq 13$.

Given a target clique $C$, how to use clique motifs attached to $C$ to characterize structures around $C$? We define $C$'s structural feature as a concatenation of 13 vectors: $[u_1^{\mathsmaller{(C)}}; u_2^{\mathsmaller{(C)}}; ..., u_{13}^{\mathsmaller{(C)}}]$. $u_i^{\mathsmaller{(C)}}$ is a vector of descriptive statistics that summarize the vectorized distribution of type-$i$ clique motifs attached to $C$. Mathematically:\vspace{-1mm}
\begin{align*}
        \hspace{3mm}u_i^{\mathsmaller{(C)}} &= \textbf{S{\scriptsize UMMARIZE}}(P_i^{\mathsmaller{(C)}})\\\vspace{-0.8mm}
    \hspace{3mm}P_i^{\mathsmaller{(C)}} &= \begin{cases}
    [\textbf{C{\scriptsize OUNT}}(C, i, \{v\}) \text{,  for } v \text{ in } C],\hspace{15mm} \text{if $1\leq i\leq 3$};\\
    [\textbf{C{\scriptsize OUNT}}(C, i, \{v1, v2\}) \text{,  for } v_1,v_2 \text{ in } C], \hspace{4.3mm}\text{if $4\leq i\leq 13$};\\
    \end{cases}
\end{align*}\vspace{-1mm}
$P_i^{\mathsmaller{(C)}}$ is a vectorized distribution in the form of an array of counts regarding $i$ and $C$. $\textbf{C{\scriptsize OUNT}}(C, i, \chi)=|\{\phi\in\Phi_i^{\mathsmaller{(C)}}|\chi\subseteq\phi\}|$. Finally, $\textbf{S{\scriptsize UMMARIZE}}(P_i^{\mathsmaller{(C)}})$ is a function that takes in a vectorized distribution $P_i^{\mathsmaller{(C)}}$ and outputs a vector of statistical descriptors. Here we simply define it to comprise of 4 basic statistical descriptors:\vspace{-1mm} $$\textbf{S{\scriptsize UMMARIZE}}(P_i^{\mathsmaller{(C)}})=[\text{\small mean}(P_i^{\mathsmaller{(C)}}), \text{\small std}(P_i^{\mathsmaller{(C)}}), \text{{\small min}}(P_i^{\mathsmaller{(C)}}), \text{\small max}(P_i^{\mathsmaller{(C)}})]\vspace{-1mm}$$
As we have 13 clique motifs, these amount to 52 structural features. \rv{On the high level, clique motifs extend the well-tested motif methods on graphs to hypergraph projections with clique structures. Compared to count features, clique motifs capture structural features in a more principled manner.}




\vspace{-0.5mm}
\section{Experiment}\label{sec:exp}
\begin{table}[]
\footnotesize
\begin{tabular}{lccp{2.9mm}p{2.9mm}p{2.9mm}p{4.2mm}}
\hline
\textbf{Dataset} &
  \textbf{$|V|$} &
  \textbf{$|\mathcal{E}|$} &
  \textbf{$\mu(\mathcal{E})$} &
  \textbf{$\sigma(\mathcal{E})$} &
  \textbf{$\bar{d}(V)$} &
  \textbf{$|\mathcal{M}|$} \\ \hline
Enron \cite{benson2018simplicial}      & 142 recipients   & 756 emails  & 3.0 & 2.0 & 16.0 & 362     \\
DBLP  \cite{benson2018simplicial}       & 319,916 authors & 197,067 papers & 3.0 & 1.7 & 1.83 & 166,571 \\
P. School \cite{benson2018simplicial}  & 242 students    & 6,352 chats   & 2.4 & 0.6 & 63.5 & 15,017  \\
H. School \cite{benson2018simplicial}  & 327 students    & 3,909 chats   & 2.3 & 0.5 & 27.8 & 3,279   \\
Foursquare \cite{young2021hypergraph}  & 2,334 restaurants  & 1,019 footprints  & 6.4 & 6.5 & 2.80 & 8,135   \\
Hosts-Virus \cite{young2021hypergraph} & 466 hosts   & 218 virus    & 5.6 & 9.0 & 2.60 & 361     \\
Directors \cite{young2021hypergraph}  & 522 directors    & 102 boards   & 5.4 & 2.2 & 1.05 & 102     \\
Crimes \cite{young2021hypergraph} & 510 victims   & 256 homicides    & 3.0 & 2.3 & 1.48 & 207    \\ \hline
\end{tabular}
\caption{\small Summary of the datasets (query split). $\mu(\mathcal{E})$ and $\sigma(\mathcal{E})$ stand for mean and std. of the distribution of the hyperedges' size. \textbf{$\bar{d}(V)$}: average degree of nodes (\textit{w.r.t.} hyperedges).  $|\mathcal{M}|$: maximal cliques.}
\label{tab:dataset}
\vspace{-8mm}
\end{table}
How well does our approach work in practice? How do we make sense of the reconstruction result? Is our approach robust to scarcity or distribution shift of training data? These questions are systematically studied in our experiments. 

We introduce settings in Sec. \ref{subsec:exp_settings}, followed by performance comparisons of all methods in Sec. \ref{subsec:exp_acc}. In Sec, \ref{subsec:exp_quality} we evaluate the reconstruction on more dimensions and check what exactly gets recovered. In Sec. \ref{subsec:exp_semi}, we evaluate reconstructions in semi-supervised setting and transferred setting --- two important extensions involving label scarcity and distribution shift, respectively. Sec. \ref{subsec:exp_abl} reports ablation studies that show the effectiveness of our clique sampler. \hide{Finally, Sec. \ref{subsec:multiedge} experiments a heuristic baseline that exploits edge multiplicities when such information becomes available.}

\noindent \textbf{Reproducibility:} Our code and data are published at \url{bit.ly/SHyRe}.  


\vspace{-1mm}
\subsection{Experimental Settings}\label{subsec:exp_settings}
\noindent\textbf{Baselines.} For comparison we adapt 7 methods originally proposed in four different task domains. They are chosen for (1) their best relevance to our task, and/or (2) their state-of-the-art status within their own task domain. The baselines and their task domains are:\vspace{-1mm}
\begin{itemize}[leftmargin=.2in]
    \item \textit{Community Detection:} Demon \cite{coscia2012demon}, CFinder \cite{palla2005uncovering}
    \item \textit{Clique Decomposition}: Max Clique \cite{bron1973algorithm}, Edge Clique Cover \cite{conte2016clique}
    \item \textit{Hyperedge Prediction:} Hyper-SAGNN \cite{zhang2019hyper}, CMM \cite{zhang2018beyond};
    \item \textit{Probabilistic Models:} Bayesian-MDL  \cite{young2021hypergraph}.
\end{itemize}\vspace{-1mm}
Since they are evaluated on our new task, adaptation is necessary, and decisions are made towards the best fairness of comparison. \rv{For example, both our method and Bayesian-MDL use maximal clique algorithm (\textit{i.e.} the Max Clique baseline) as a preprocessing step. Our evaluation ensures that all three of them use the same implementation as introduced in \cite{bron1973algorithm}.}  See Appendix \ref{sec:more_exp} for more details on selection criteria and adaptation.

\noindent\textbf{Data Preparation.} We use 8 real-world datasets containing different high-order relationships: email correspondence, paper coauthorship, social interactions, biological groups, shared restaurants \textit{etc}. They feature significantly varying number of nodes, hyperedges, distribution of hyperedge sizes, \textit{etc.} summarized in Table \ref{tab:dataset}.

To generate a training set and a query set, we follow two common standards to split the collection of hyperedges in each dataset: (1) For datasets that come in natural segments, such as DBLP and Enron whose hyperedges are timestamped, we follow their segments so that training and query contain two disjoint and roughly equal-sized sets of hyperedges. \rv{For DBLP, we construct $\mathcal{H}_0$ from year 2011 and $\mathcal{H}_1$ from year 2010; for Enron, we use 02/27/2001, 23:59 as a median timestamp to split all emails into $\mathcal{H}_0$ (first half) and $\mathcal{H}_1$ (second half).} (2) For all the other datasets that lack natural segments, we randomly split the set of hyperedges into halves. To enforce inductiveness, we randomly re-index node IDs in each split. \rv{Finally, we project $\mathcal{H}_0$ and $\mathcal{H}_1$ to get $G_0$ and $G_1$ respectively.}

Beyond the two common standards, we conduct semi-supervised reconstruction and transfer reconstruction in Sec. \ref{subsec:exp_semi}. 


\noindent \textbf{Training Configuration.}
For models requiring back propagation, we use cross entropy loss and optimize using Adam for $2000$ epochs and learning rate $0.0001$. Those with randomized modules are repeated 10 times with different seeds. See \ref{sec:more_exp} for more tuning details.

\vspace{-3mm}
\subsection{Quality of Reconstruction}\label{subsec:exp_acc}

\begin{table*}[]
\begin{tabular}{lcccccccc}
\hline
 &
  \textbf{DBLP} &
  \textbf{Enron} &
  \textbf{P.School} &
  \textbf{H.School} &
  \textbf{Foursquare} &
  \textbf{Hosts-Virus} &
  \textbf{Directors} &
  \textbf{Crimes} \\ \hline

CFinder \cite{palla2005uncovering}         & 11.35 & 0.45    & 0.00 & 0.00 & 0.39  & 5.02  & 41.18  & 6.86  \\
Demon \cite{coscia2012demon}          & -          & 2.35 & 0.09 & 2.97 & 16.51& 7.28  & 90.48  & 63.81 \\
Maximal Clique \cite{bron1973algorithm}      & 79.13 & 4.19    & 0.09 & 2.38 & 9.62  & 22.41 & \textbf{100.0} & 78.76 \\
Clique Covering \cite{conte2016clique} & 73.15 & 6.61    & 1.95 & 6.89 & \textbf{79.89} & 41.00 & \textbf{100.0} & 75.78 \\
Hyper-SAGNN \cite{zhang2019hyper}     & 0.12$\pm$0.01  &    0.19$\pm$0.17        & 12.13$\pm$0.33       &  8.89$\pm$0.15      &   0.01$\pm$0.01       &  7.35$\pm$0.48       &  1.94$\pm$1.12       & 0.86$\pm$ 0.17        \\
CMM  \cite{zhang2018beyond}           & 0.11$\pm$0.04       & 0.48$\pm$0.06        & 14.26$\pm$0.92     & 4.27$\pm$0.46    & 0.00$\pm$0.00    &  6.44$\pm$ 0.82  & 2.55$\pm$0.62      & 0.57$\pm$0.29      \\
Bayesian-MDL \cite{young2021hypergraph} &
  73.08$\pm$0.00 &
  4.57$\pm$0.07 &
  0.18$\pm$0.01 &
  3.58$\pm$0.03 &
  69.93$\pm$0.59 &
  40.24$\pm$0.12 &
  \textbf{100.0$\pm$0.00} &
  74.91$\pm$0.11 \\\hdashline
\textbf{\method-count} &
  \textbf{81.18$\pm$0.02} &
  13.50$\pm$0.32 &
  42.60$\pm$0.61 &
  \textbf{54.56$\pm$0.10} &
  74.56$\pm$0.32 &
  \textbf{48.85$\pm$0.11} &
  \textbf{100.0$\pm$0.00} &
  79.18$\pm$0.42 \\
\textbf{\method-motif} &
  \textbf{81.19$\pm$0.02} &
  \textbf{16.02$\pm$0.35} &
  \textbf{43.06$\pm$0.77} &
  54.39$\pm$0.25 &
  71.88$\pm$0.28 &
  45.16$\pm$0.55 &
  \textbf{100.0$\pm$0.00} &
  \textbf{79.27$\pm$0.40} \\\hline
\end{tabular}
\caption{Comparison of different methods on hypergraph reconstruction, performance measured in Jaccard Similarity (\%). Standard deviation is dropped where the method is deterministic or randomization has no effect on performance. ``-'' means the algorithm did not stop in 72 hours.}
\label{tab:results}\vspace{-5mm}
\end{table*}
We name our approach \textbf{\method} (\textbf{S}upervised \textbf{Hy}pergraph \textbf{Re}construction). Table \ref{tab:results} shows the quality of reconstruction measured in Jaccard Score (see Sec. \ref{sec:def}).  The two variants of \method strongly outperform all baselines on most datasets (7/8). The improvement is most significant on hard datasets such as P. School, H.School and Enron. We attribute the success to our reconstruction framework that makes the best out of supervised signals: we derive insights from our topological analysis, and find ways to efficiently probe high-quality candidates for hyperedges and for designing good classifiers. 

The two variants are close to each other by performance. \method-motif very marginally wins if we count the number of datasets on which one variant achieves the best score. We further study the two variants and found that they both capture a strong feature (more in \ref{sec:exp}). Among the baselines, Clique Covering and Bayesian-MDL work relatively well. Still, the ``principle of parsimony'' \cite{young2021hypergraph} suffers on dense hypergraphs if we interpret the result according to Table \ref{tab:dataset}. 

\rv{Fig.\ref{fig:partitioned_performance} further visualizes fine-grained performance measured by partitioned errors as defined in Def. \label{def:errors}. Further explanations are included in the caption. We can see that \method variants
significantly reduce more Error I and II than the baselines.}

\vspace{-3mm}
\subsection{Further Evaluations of Reconstruction}\label{subsec:exp_quality}
Beyond Jaccard score, we seek to further understand \textbf{(a)} whether our reconstructions preserve important properties of the original hypergraphs, \textbf{(b)} the running time of our reconstructions, and, more curiously, \textbf{(c)} what exactly gets recovered. To this end, we study the property space, the asymptotic running time, and an actual visualization of the reconstructions.

\vspace{1mm}
\noindent\textbf{Property Space.} The procedure works as follows: we first characterize the property space of the reconstruction using the middle four columns of Table \ref{tab:dataset}: $[|\mathcal{E}|,\mu(\mathcal{E}),\sigma(\mathcal{E}), \bar{d}(V)]$. $|V|$ is excluded from the property vector as it is known from the input. Then for each (dataset, method) combination, we analyze the reconstruction and obtain a unique property vector. We use PCA to project all (normalized) property vectors into 2D space, visualized in Fig.\ref{fig:vis_stats}.

In Fig.\ref{fig:vis_stats}, colors encode datasets, and marker styles encode methods. Compared with baselines, \method variants ( \Circle and $\times$) produce reconstructions more similar to the ground truth ($\blacksquare$). The reasons are two-fold: (1) \method variants usually achieve better accuracy, which encourages a more aligned property space; (2) as a bonus of our greedy Alg. \ref{algo:sampler}, in each iteration it tends to look for a cell from a different column. This is because cells in the same column has diminishing returns due to overlapping of $\mathcal{E}_{n,k}$ with same $k$, whereas cells in different columns remain unaffected as they have hyperedges of different sizes. The inclination of having diverse hyperedge sizes reduces the chance of a skewed distribution.

We also observe that markers of the same colors are generally clustered, meaning that most baselines work to some extent despite low accuracy sometimes. Fig.\ref{fig:vis_stats} also embeds a historgram showing the size distribution of the reconstructed hyperedges on DBLP. We see the distribution obtained by \method aligns decently with the ground truth, especially with large hyperedges. Some errors are made with sizes 1 and 2, which are mostly the Fig.\ref{fig:error_diagram} nested cases.

\begin{figure}[t]
    \centering\vspace{-5mm}
    \includegraphics[scale=0.38]{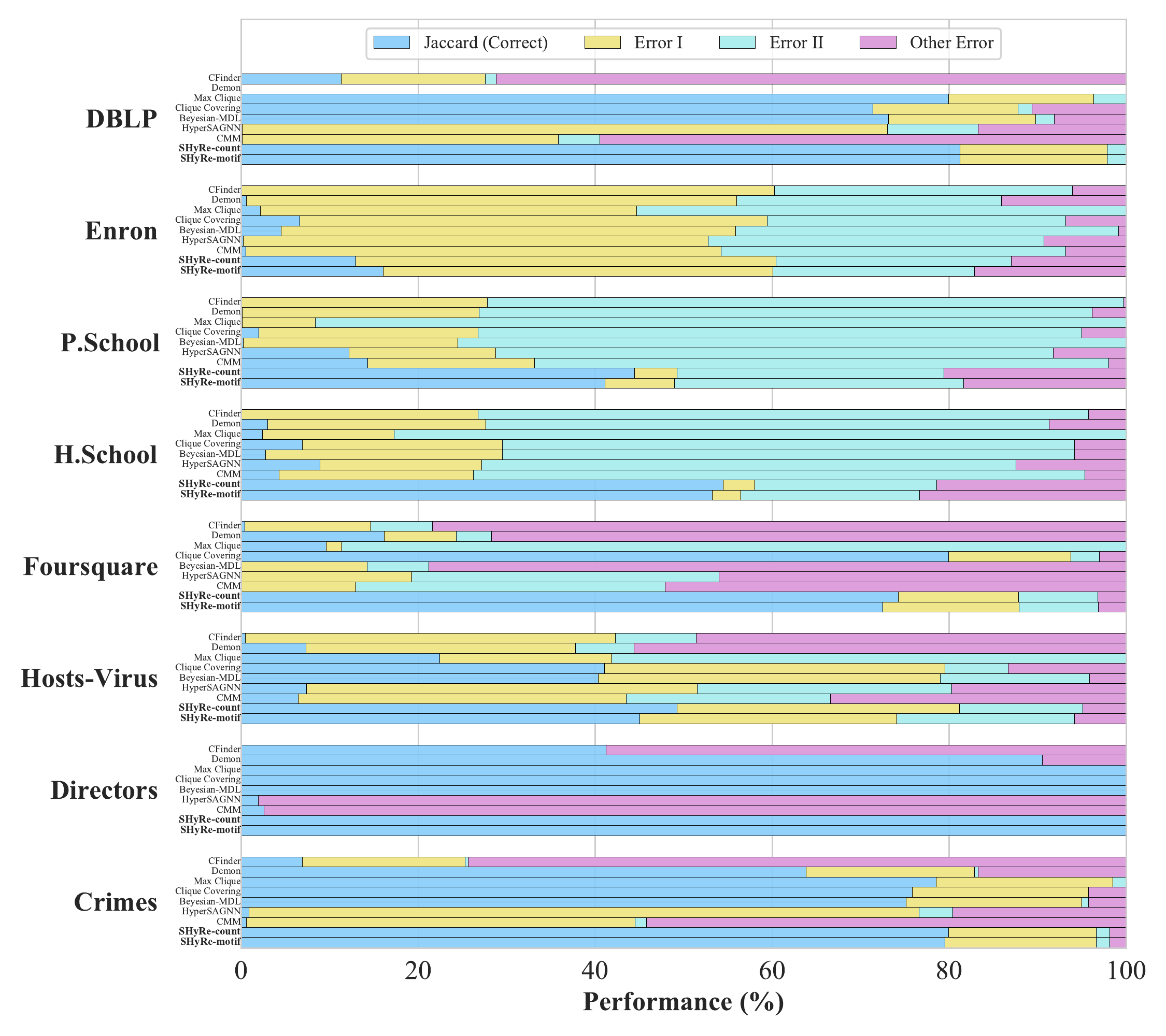}\vspace{-4mm}
    \caption{\rv{Partitioned Performance of all methods on all datasets. Recall that the Error I and II are mistakes made by Max Clique (Def.\ref{def:errors}). Other methods may make mistakes that Max Clique does not make, which are counted as ``Other Error''. We can see that \method reduce more Error I and II than other baselines.}}
    \label{fig:result_split}\vspace{-2mm}
\end{figure}
\begin{figure}
    \centering\vspace{-2.8mm}
    \includegraphics[scale=0.30]{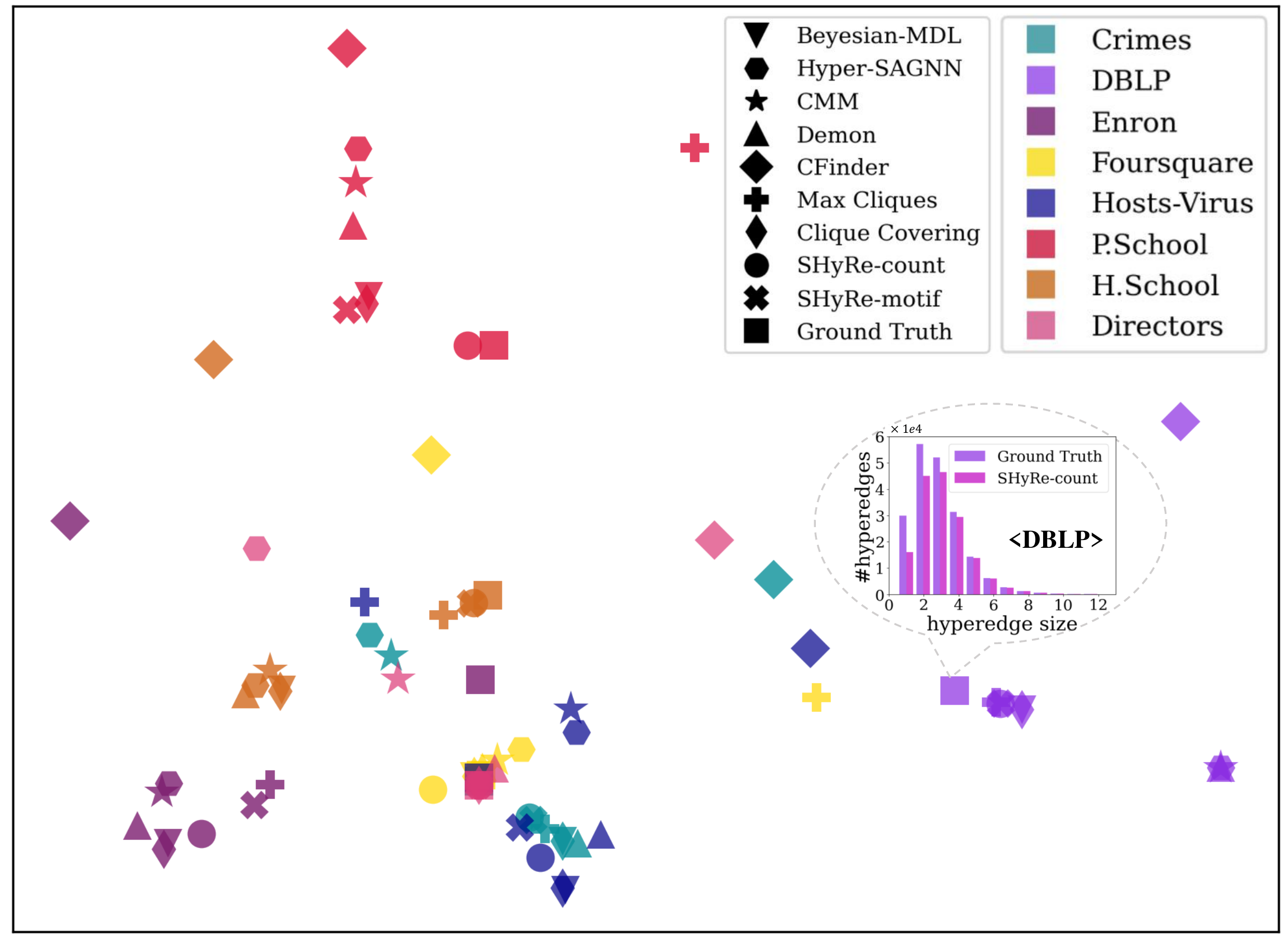}\vspace{-4mm}
    \caption{\small {2D embeddings of statistical properties of reconstructed hypergraphs. Different colors correspond to different datasets; different markers correspond to different methods. Notice that reconstructions of \method are closest to \rv{ground truth ($\blacksquare$)} on most datasets.}}
    \label{fig:vis_stats}\vspace{-2mm}
\end{figure}

\begin{figure}
    \centering\vspace{-4mm}
    \includegraphics[scale=0.42]{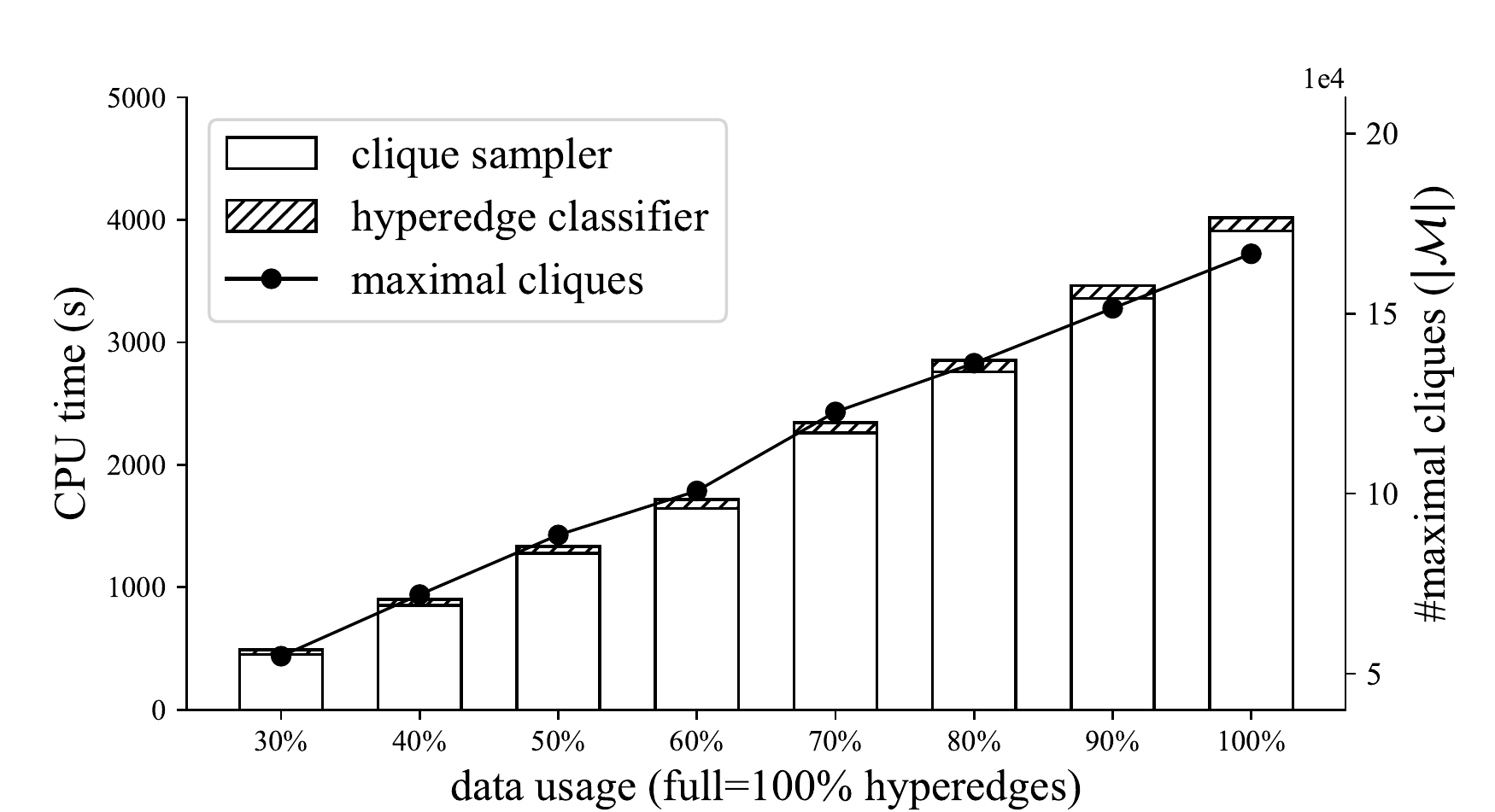}\vspace{-2mm}
    \caption{\small {\yb{Asymptotic running time analysis of \method-motif on DBLP. The bar plot is aligned with the left y-axis, and the line plot with the right.  Notice that both the total CPU time and the number of maximal cliques are roughly linear to the data usage (size).}}}
    \label{fig:vis_runtime}
    \vspace{-4mm}
\end{figure}

\noindent\textbf{Running Time.} We claim in Sec. \ref{sec:framework} that the clique sampler's complexity is close to $O(|\mathcal{M}|)+O(|\mathcal{E}|)=O(|\mathcal{M}|)$ in practice. Here this claim is substantiated with an asymptotic running time analysis. Both the clique sampler (Step 1, 2 in Fig.\ref{fig:framework}) and the hyperedge classifier  (Step 3, 4 in Fig.\ref{fig:framework}) are tested. For $p\in[30, 100]$, We randomly sample $p\%$ hyperedges from DBLP and record the CPU time for running both modules of \method-motif. The result is plot in Fig.\ref{fig:vis_runtime}. It shows that both the total CPU time and the number of maximal cliques are roughly linear to the data usage (size), which verifies our claim. \rv{We provide additional statistics all methods' time complexity in Fig.\ref{fig:time} Appendix.}

\begin{figure}[H]
    \centering
    \vspace{-3.5mm}
    \includegraphics[scale=0.25]{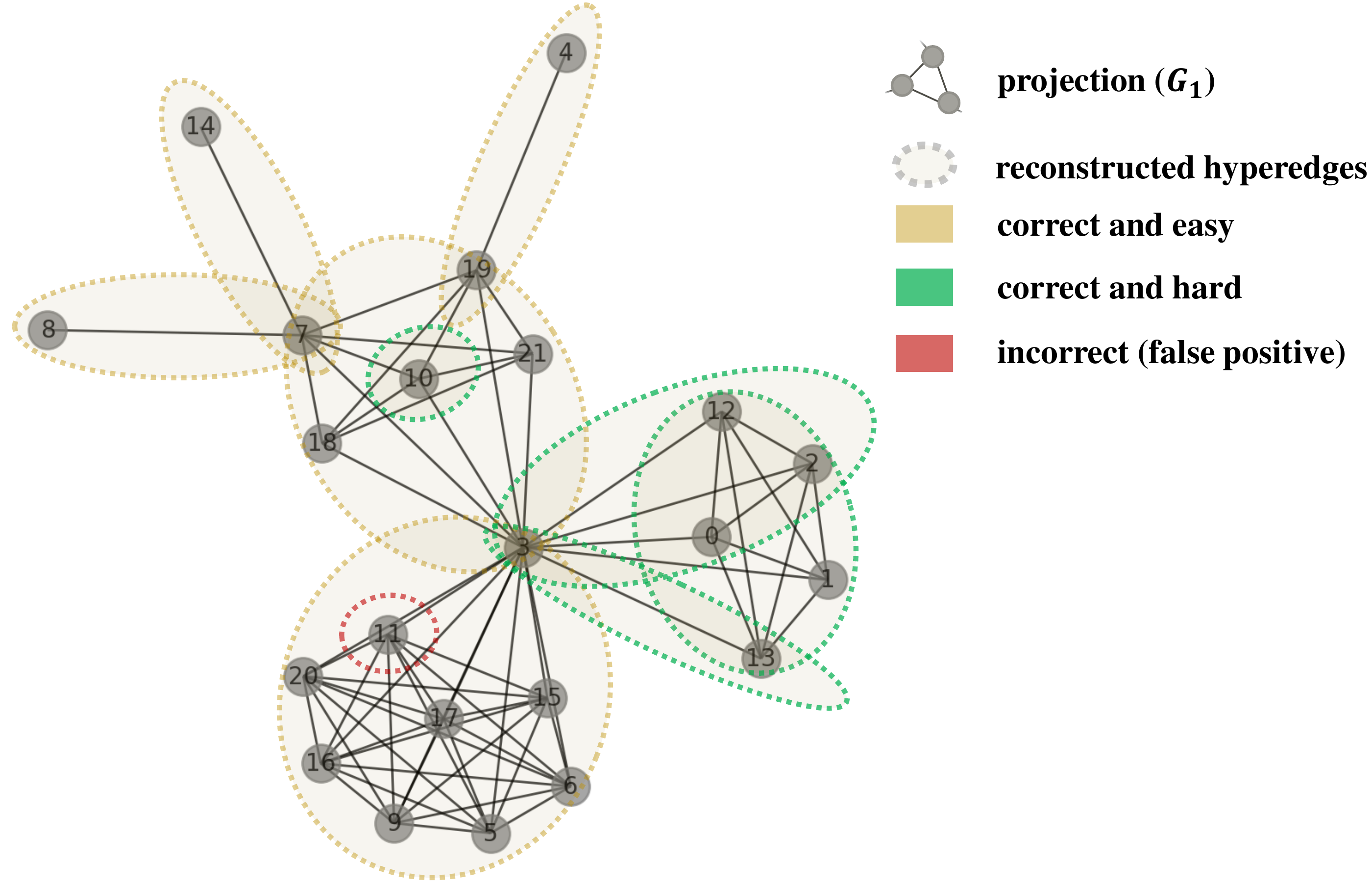}
    \caption{\small A small part of the hypergraph reconstructed by \method-motif on DBLP dataset. Projection $G_1$ is drawn in black. The shaded ellipsis are the reconstructed hyperedges. Those with green dashed edges are difficult to reconstruct in the absence of the training data. Notice that $\{3, 12, 2, 0, 13, 1\}$ is a maximal clique (in this local graph).}\label{fig:vis_graph}
    \vspace{-3mm}
\end{figure}

\noindent\textbf{Visualizing the Reconstruction.} Fig.\ref{fig:vis_graph} visualizes a portion of the actual reconstructed hypergraph by \method-count on DBLP dataset. We include more explanation and analysis in the caption.

\vspace{-1mm}
\subsection{Label Scarcity and Distribution Shift}\label{subsec:exp_semi}
So far, we have assumed that our training split consists of a comparable number of hyperedges as the query split.  However, sometimes we may only have access to a small subset of hyperedges or a different but similar dataset. Will the scarcity of training data or the distribution shift become a problem? We conduct mini-studies to find out more about reconstruction based on semi-supervised learning a well as transfer learning.

We choose three datasets of different reconstruction difficulties: DBLP, Hosts-Virus, and Enron. For each, we randomly drop 80\% of the hyperedges in the training split.  Our framework is trained and tuned the same way as in the main experiment. 


Table \ref{tab:semi} shows the result.  We can see the accuracy of the reconstruction is negatively influenced on all datasets. However, \method on 20\% data still outperforms the best baseline on full data, meaning that it remains well functioning in a semi-supervised setting.  Comparing column-wise, the influence is negligible on easy datasets, small on moderately hard datasets, and large on very hard datasets.

\begin{figure}
    \centering\vspace{-2mm}
    \includegraphics[scale=0.35]{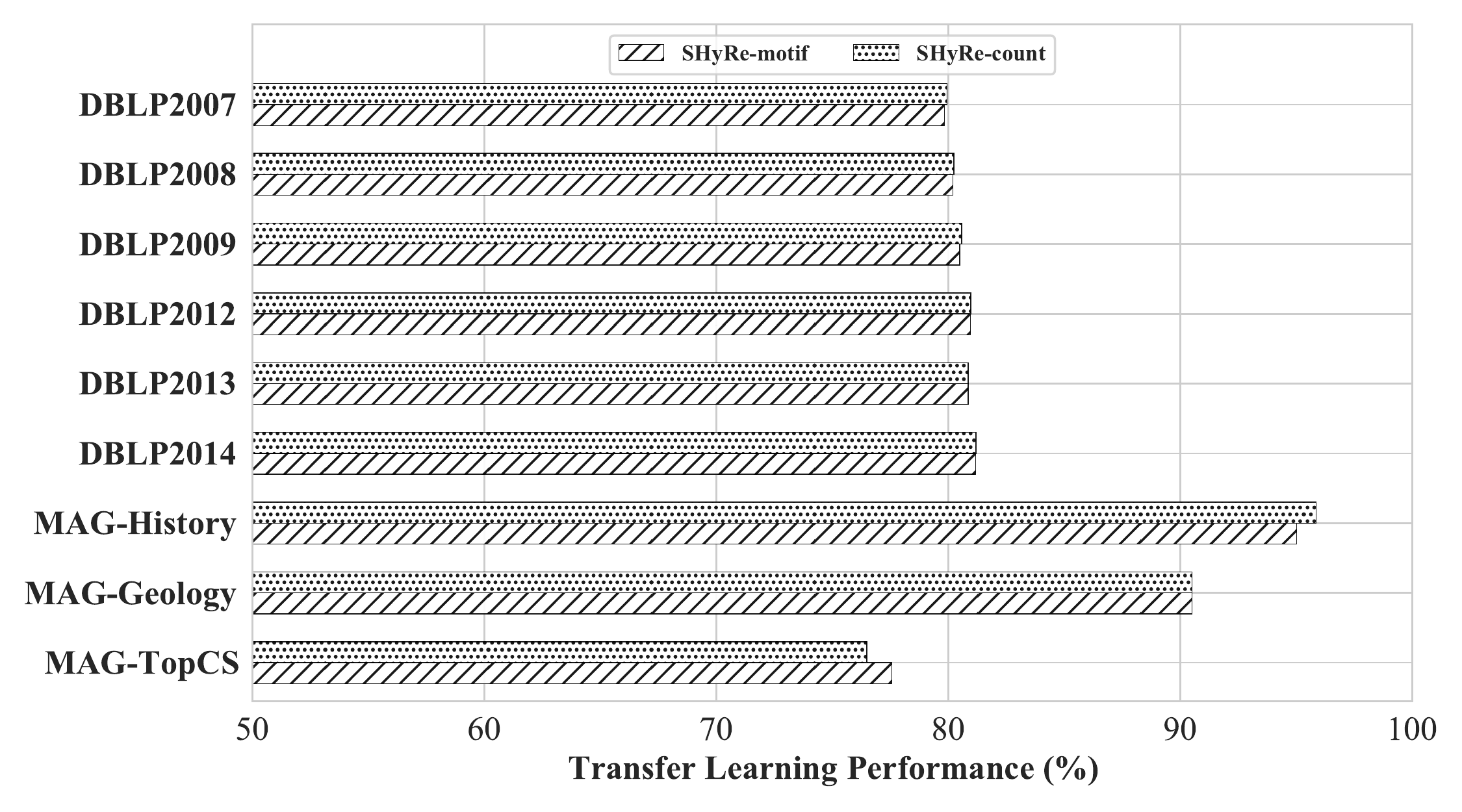}\vspace{-4mm}
    \caption{\small \rv{Performance of transfer learning: trained on DBLP2011 and tested on various other coauthorship datasets.}}
    \label{fig:transfer}\vspace{-4mm}
\end{figure}

\rv{
For transfer learning, we train on DBLP 2011, and test \method's performance on various other DBLP slices as well as Microsoft Academic Graphs (MAG), a different bibliographical database similar to DBLP. Fig.\ref{fig:transfer} shows the result. We can see that \method remains relatively robust to the distribution shift.}

\begin{table}[]
\small
\begin{tabular}{lccc}
\hline
\textbf{}      & \textbf{DBLP}  & \textbf{Hosts-Virus} &\textbf{Enron}   \\ \hline
Best Baseline (full) & 79.13          & 41.00           &6.61                 \\
\method-motif (full) & 81.19$\pm$0.02 & 45.16$\pm$0.55 &16.02$\pm$0.35        \\
\method-count (20\%) & 81.17$\pm$0.01 & 44.02$\pm$0.39  &6.43$\pm$0.18        \\
\method-motif (20\%) & 81.17$\pm$0.01 & 44.48$\pm$0.21 & 10.56$\pm$0.92       \\ 
\hline
\end{tabular}
\caption{\small Performance of semi-supervised reconstruction using 20\% of the training hyperedges, measured in Jaccard Similarity.}\vspace{-3mm}
\label{tab:semi}\vspace{-5mm}
\end{table}

\vspace{-1mm}
\subsection{Ablation Study on Clique Sampler}\label{subsec:exp_abl}
In Sec. \ref{subsec:sampler} we extensively discuss how to optimize the clique sampler. One might argue that the optimization appears complex: can we adopt a simpler heuristic for sampling which works independently of the notion of $r_{n,k}$'s? We investigate this via an ablation study.

We test three sampling heuristics that might replace the clique sampler. 1.\textbf{ ``random''}: we randomly sample $\beta$ cliques from the projection as candidates. While it is hard to achieve strict uniformness \cite{khorvash_2009}, we approximate this by growing a clique from a random node and stopping the growth when the clique reaches a random size; 2.\textbf{``small''}: we randomly sample $\beta$ cliques of sizes 1 and 2 (\textit{i.e.} nodes and edges); 3.\textbf{``head \& tail''}: we randomly sample $\beta$ cliques from all cliques  of sizes 1 and 2 as well as maximal cliques.

\begin{figure}
    \centering
    \includegraphics[scale=0.24]{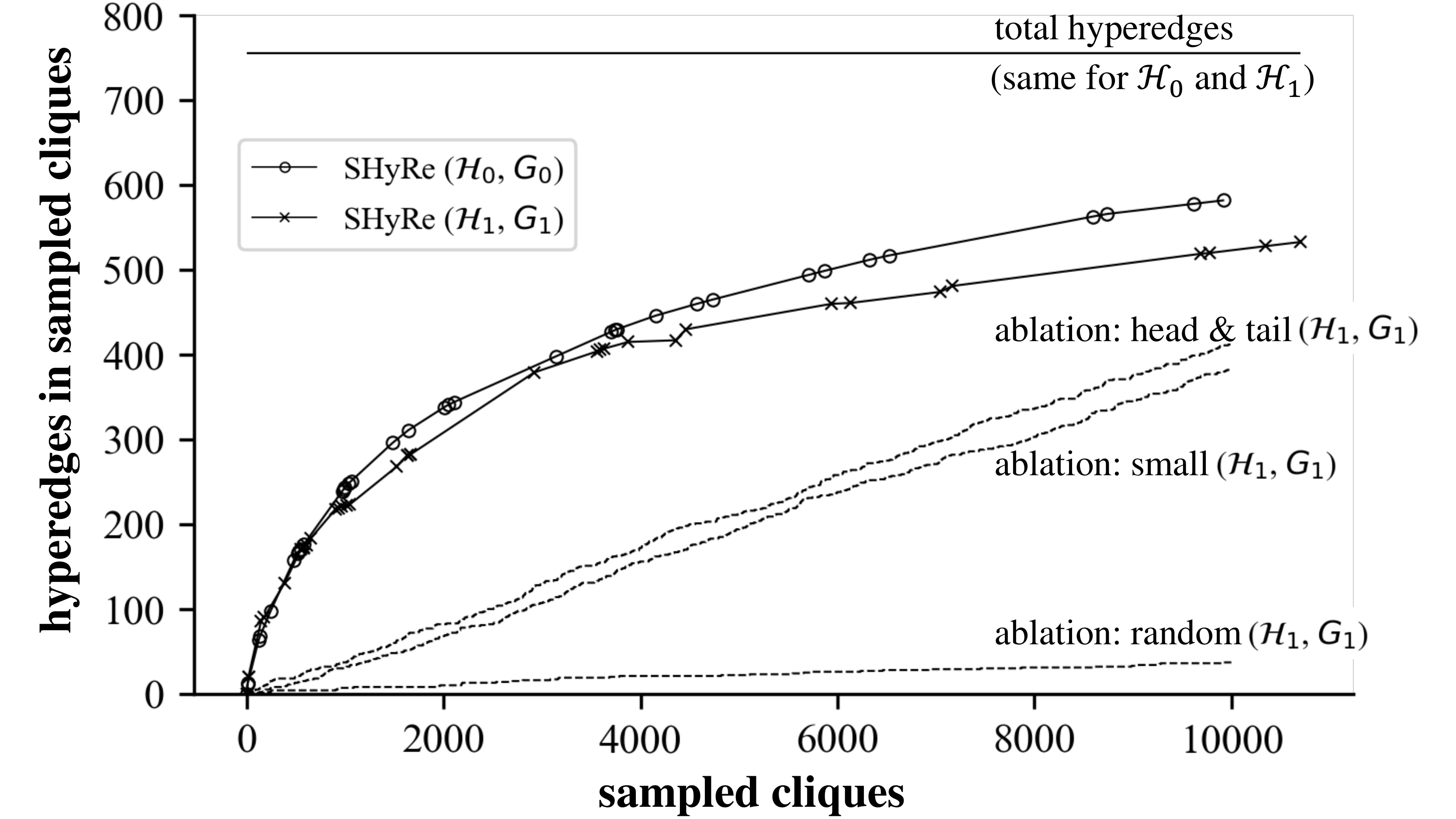}\vspace{-2.3mm}
    \caption{\small Our clique sampler and ablation studies. Each marker is an iteration in Alg. \ref{algo:sampler}. The alignment between the training curve and the query curve shows that our clique sampler generalizes well.} 
    \label{fig:exp_sampler}\vspace{-3mm}
\end{figure}

Fig.\ref{fig:exp_sampler} compares the efficacy in the sampling stage on Enron dataset. It shows that our clique sampler significantly outperforms all heuristics and so it cannot be replaced. Also, the the great alignment between the training curve and query curve means our clique sampler generalizes well. We further report reconstruction performance on 3 datasets in Table \ref{tab:exp_ablation}, which also confirms this point. 



\vspace{-4mm}
\section{Additional Related Work}\label{sec:survey}

There are three lines of work pertinent to the reconstruction task.

\noindent\textbf{Edge Clique Covering} is to find a minimal set of cliques that cover all the graph's edges. The clique-based projection makes this task highly relevant. \cite{erdos1966representation} proves any graph can be covered by at most $[\frac{|V|^2}{4}] $ cliques. \cite{conte2016clique} finds a fast heuristic for approximating the solution. \cite{young2021hypergraph} creates a probabilistic framework to redefine and solve the task. However, this line of work shares the ``principle of parsimony'', which is often impractical in real-world datasets.

\noindent\textbf{Hyperedge Prediction} is to identify missing hyperedges of an incomplete hypergraph from a pool of given candidates. Existing work focuses on characterize a node set's structural features. The methods span proximity measures \cite{benson2018simplicial}, deep learning \cite{li2020distance,zhang2019hyper}, and matrix factorization \cite{zhang2018beyond}. Despite the relevance, the task has a very different setting and focus from ours as mentioned in Sec. \ref{sec:intro}.

\noindent\textbf{Community Detection} finds node clusters within which edge density is much higher than the outside. Existing work roughly comes in two categories per the community type: disjoint \cite{que2015scalable,traag2019louvain}, and overlapping \cite{coscia2012demon,palla2005uncovering}. As mentioned, however, the notion of ``relative density'' is not compatible with our focus on cliques. 

\vspace{-1.5mm}
\section{Conclusion}
We propose the supervised hypergraph reconstruction task to effectively understand and compensate for the loss of high-order information common in graph data analysis. Our well-motivated reconstruction framework consists of a clique sampler and a hyperedge classifier. Its success is substantiated by extensive experiments. For future work, our setting can be extended in many meaningful directions. For example, how can we improve the reconstruction if we have node attributes or edge multiplicity? What if a hyperedge is projected not into a clique but into a ``star'' or a ``line''? These studies will substantially benefit graph analysis and scientific discovery.


\vspace{-1mm}
\bibliographystyle{ACM-Reference-Format}
\bibliography{references.bib}


\appendix
\clearpage\newpage
\vspace{-2mm}
\section{Proofs}\vspace{-1mm}
\subsection{Proof of Theorem \ref{theorem:sperner_conformal}}
{\parindent0pt
\begin{proof}
From Def. \ref{def:conformal}, it suffices to show that for every $C$ that is not a maximal clique, $C$ is not in $\mathcal{E}$. This holds because in that case $C$ has to be the proper subset of some maximal clique $C'\in \mathcal{E}$, but since $H$ is Sperner, $C$ cannot be a hyperedge.
\end{proof}}

\vspace{-4mm}\subsection{Proof of Theorem \ref{theorem:triangle}}\label{subsec:triangle}\vspace{-1mm}
{\parindent0pt\begin{proof}
The ``if'' direction: Suppose that $\mathcal{H}$ is not conformal. 
According to Def. \ref{def:conformal}, we know that there exists a maximal clique $C\notin \mathcal{E}$. Clearly for every $C$ with $|C|\leq2$, $|C|\in \mathcal{E}$. For a $C\notin \mathcal{E}$ and $|C|\geq 3$, pick any two nodes $v_1, v_2\in C$. Because they are connected, $v_1, v_2$ must be in some hyperedge $E_i$. Now pick a third node $v_3\notin E_i$. Likewise, there exists some different $E_j$ such that $v_1, v_3\in E_2$, and some different  $E_q$ such that $v_2, v_3\in E_3$. Notice that $E_j\neq E_q$ because otherwise the three nodes would be in the same hyperedge. Now we have $\{v_1, v_2, v_3\}\subseteq (E_i \cap E_j) \cup (E_j \cap E_q) \cup (E_q \cap E_i)$. Because $\{v_1, v_2, v_3\}$ is not in the same hyperedge, $(E_i \cap E_j) \cup (E_j \cap E_q) \cup (E_q \cap E_i)$ is also not in the same hyperedge.

The ``only if'' direction: Because every two of the three intersections share a common hyperedge, their union is a clique. The clique must be contained by some hyperedge, because otherwise the maximal clique containing the clique is not contained by any hyperedge.
\end{proof}}

\vspace{-2mm}
Alternatively, there is a less intuitive proof that builds upon results from existing work in a detour: It can be proved that $\mathcal{H}$ being conformal is equivalent to its dual $\mathcal{H}'$ being Helly \cite{berge1973graphs}. According to an equivalence to Helly property mentioned in \cite{berge1975generalization}, for every set $A$ of 3 nodes in $\mathcal{H}'$, the intersection of the edges $E_i$ with $|E_i\cap A|\geq k$ is non-empty. Upon a dual transformation, this result can be translated into the statement of Theorem \ref{theorem:sperner_conformal}. We refer the interested readers to the original text.

\vspace{-1mm}
\subsection{Proof of Theorem \ref{theorem:count}} \label{subsec:count}

Given a set $X=\{1, 2, ..., m\}$ and a hypergraph $\mathcal{H}=(V, \{E_1, ... E_m\})$, we define $f:2^X\rightarrow2^V$ to be: \vspace{-1mm}
$$f(S)=\left( \bigcap_{i\in S} E_i\right) \bigcap\left(\bigcap_{i\in X\backslash S}\bar{E_i}\right),\; S\subseteq X$$\vspace{-1mm}
where $\bar{E_i}=V\backslash E_i$.

{\parindent0pt
\begin{lemma}\label{lemma:partition}
$\{f(S)|S\in 2^X\}$ is a partition of $V$.
\end{lemma}\vspace{-3mm}
\begin{proof}\vspace{-2mm}
Clearly $f(S_i) \cap f(S_j)=\emptyset$ for all $S_i \neq S_j$, so elements in $\{f(S)|S\in 2^X\}$ are disjoint. Meanwhile, for every node $v\in V$, we can construct a $S=\{i|v\in E_i\}$ so that $v\in f(S)$. Therefore, the union of all elements in $\{f(S)|S\in 2^X\}$ spans $V$. 
\end{proof}
\vspace{-3mm}
Because Lemma \ref{lemma:partition} holds, for any $v\in V$ we can define the reverse function $f^{-1}(v)=S\Leftrightarrow v\in f(S)$. Here $f^{-1}$ is a signature function that represents a node in $V$ by a subset of $X$, whose physical meaning is the intersection of hyperedges in $\mathcal{H}$.

\begin{lemma}\label{lemma:intersect}
If $S_1 \cap S_2 \neq \emptyset, S_1, S_2 \subseteq X$, then for every $v_1 \in f(S_1)$ and $v_ 2\in f(S_2)$, $(v_1, v_2)$ is an edge in $H$'s projection $G$. Reversely, if $(v_1, v_2)$ is an edge in $G$, $f^{-1}(v_1)\cap f^{-1}(v_2)\neq \emptyset$.
\end{lemma}

\vspace{-3mm}
\begin{proof}
According to the definition of $f(S)$, $\forall i\in S_1 \cap S_2,\; v_1, v_2 \in E_i$. Appearing in the same hyperedge means that they are connected in $G$, so the first part is proved. If $(v_1, v_2)$ is an edge in $G$, there exists an $E_i$ that contains both nodes, so $f^{-1}(v_1)\cap f^{-1}(v_2) \supseteq \{i\}\neq\emptyset$.
\end{proof}}

\vspace{-2mm}
\noindent An \textit{intersecting family} $\mathcal{F}$ is a set of non-empty sets with non-empty pairwise intersection, \textit{i.e.} $S_i\cap S_j \neq \emptyset,\;\; \forall S_i, S_j \in \mathcal{F}$. Given a set $X$, a \textit{maximal intersecting family of subsets}, is an intersecting family of set $\mathcal{F}$ that satisfies two additional conditions: (1) Each element of $\mathcal{F}$ is a subset of $X$; (2) No other subset of $X$ can be added to $\mathcal{F}$.

{\parindent0pt\begin{lemma}\label{lemma:mapping}
Given a a hypergraph $\mathcal{H}=(V, \{E_1, E_2, ... E_m\})$, its projection $G$, and  $X=\{1, 2, ..., m\}$, the two statements below are true:
\begin{itemize}[leftmargin=3mm]
    \item If a node set $C \subseteq V$ is a maximal clique in $G$, then $\{f^{-1}(v)|v\in C\}$ is a maximal intersecting family of subsets of $X$.
    \item  Reversely, if $\mathcal{F}$ is a maximal intersecting family of subsets of $X$, then $\cup_{S\in\mathcal{F}}f(S)$ is a maximal clique in $G$.
\end{itemize}
\end{lemma}}

{\parindent0pt
\begin{proof}\vspace{-1mm}
For the first statement, clearly $\forall v\in V, \;f^{-1}(v) \subseteq X$. Because $C$ is a clique, every pair of nodes in $C$ is an edge in $G$. According to Lemma \ref{lemma:intersect}, $\forall v_1, v_2 \in C, \;f^{-1}(v_1)\cap f^{-1}(v_2) \neq\emptyset$. Finally, because $C$ is maximal, there does not exist a node $v\in V$ that can be added to $C$. Equivalently there does not exist a $S=f^{-1}(v)$ that can be added to $f^{-1}(C)$. Therefore, $f^{-1}(C)$ is maximal.

For the second statement, because $\mathcal{F}$ is an intersecting family, $\forall S_1, S_2 \subseteq \mathcal{F}, \;S_1 \cap S_2 \neq \emptyset$. According to Lemma \ref{lemma:intersect}, $\forall v_1, v_2\in f(\mathcal{F})$, $(v_1, v_2)$ is an edge in $G$. Therefore, $f(\mathcal{F})$ is a clique. Also, no other node $v$ can be added to $f(\mathcal{F})$. Otherwise, $f^{-1}(v)\cup \mathcal{F}$ is still an intersecting family while $f^{-1}(v)$ is not in $\mathcal{F}$, which makes $\mathcal{F}$ strictly larger --- a contradiction. Therefore, $\cup_{S\in\mathcal{F}}f(S)$ is a maximal clique.

\end{proof}}

\vspace{-4mm}
\noindent Lemma \ref{lemma:mapping} shows there is a bijective mapping between a maximal clique and a maximal intersecting family of subsets. Given $\mathcal{H}$, $G$ and $X$, Counting the former is equivalent to counting the latter. The result is denoted as $\lambda(m)$ in the main text. Lemma 2.1 of \cite{brouwer2013counting} gives an lower bound: $\lambda(m)\geq2^{{m-1 \choose [m/2]-1}}$.

\vspace{-1mm}
\subsection{Proof of Theorem \ref{theorem:sampler}}\label{subsec:proof_sampler}
We start with some definitions.  A \textit{random finite set}, or RFS, is defined as a random variable whose value is a finite set. Given a RFS $A$, we use $\mathcal{S}(A)$ to denote $A$'s sample space; for a set $a\in \mathcal{S}(A)$ we use $P_A(a)$ to denote the probability that $A$ takes on value $a$. One way to generate a RFS is by defining the \textit{set sampling} operator $\odot$ on two operands $r$ and $X$, where $r\in[0,1]$ and $X$ is a finite set: $r\odot X$ is a RFS obtained by uniformly sampling elements from $X$ at sampling rate $r$, \textit{i.e.} each element $x\in X$ has probability $r$ to be kept. Also, notice that the finite set $X$ itself can also be viewed as a RFS with only one possible value to be taken.  Now, we generalize two operations, union and difference, to RFS as the following:
\begin{itemize}
    \item \textbf{Union} $A\cup B$: 
    \begin{align*}\vspace{-1mm}
        \mathcal{S}(A\cup B)&=\{x|x=a\cup b, a\in A, b\in B\}\\
        P_{A \cup B}(x)&=\sum_{x=a\cup b, a\in A, b\in B}P_A(a)P_B(b)
    \end{align*}\vspace{-3mm}
    \item \textbf{Difference} $A\backslash B$: 
    \begin{align*}
        \mathcal{S}(A\backslash B)&=\{x|x=a\backslash b, a\in A, b\in B\}\\
        P_{A \cup B}(x)&=\sum_{x=a\backslash b, a\in A, b\in B}P_A(a)P_B(b)\vspace{-1mm}
    \end{align*}
\end{itemize}
The \textit{Expectation of the Cardinality} of a RFS is denoted by $\eoc$ such that  $\eoc[A]=\mathbb{E}_{a\in\mathcal{S}(A)} |a|$. With these ready, we have the following propositions that hold true for RFS $A$ and $B$:
\begin{enumerate}[label=(\roman*)]
    \item $\eoc[A\cup B] = \eoc[B\cup A]$
    \item $\eoc[A\cup B] = \eoc[A \backslash B] + \eoc[B]$;
    \item $\eoc[A\cup B] \geq \eoc[A]$, $\eoc[A\cup B] \geq \eoc[B]$;
    \item $\eoc[(r\odot X)\backslash Y]=r|X\backslash Y|=\eoc[X\backslash Y]$;\;\;  ($X$, $Y$ are both set)
\end{enumerate}


{\parindent0pt
\begin{lemma}\label{lemma:reduce_gap}
\hspace{-0.5mm}At iteration $(i+1)$ when Algo. \ref{algo:sampler} samples a cell $(n,k)$ (l-\ref{line:cell}), it reduces the gap between $q^*$ and the expected number of hyperedges it already collects, $q_i$, by a fraction of at least $\frac{r_{n,k}|Q_{n,k}|}{\beta}$: \vspace{-1mm}
$$\frac{q^*-q_{i+1}}{q^*-q_{i}}\leq 1-\frac{r_{i+1}|Q_{i+1}|}{\beta}$$
\end{lemma}\vspace{-5mm}
\begin{proof}\vspace{-2mm}
\begin{align*}
    &\;q^*-q_{i}\\
    =&\;\eoc[\cup_{j=1}^{z} r^*_j\odot\mathcal{E}_j] - \eoc[\cup_{j=1}^i r_j\odot\mathcal{E}_j]\cim{Thm. \ref{theorem:sampler} setup}\\
    \leq&\; \eoc[(\cup_{j=1}^{z} r^*_j\odot\mathcal{E}_j) \cup (\cup_{j=1}^i r_j\odot\mathcal{E}_j)]-\eoc[\cup_{j=1}^i r_j\odot\mathcal{E}_j]\cim{Prop.iii}\\
    =&\; \eoc[(\cup_{j=1}^{z} r^*_j\odot\mathcal{E}_j)\backslash(\cup_{j=1}^i r_j\odot\mathcal{E}_j)]\cim{Prop.ii}\\
    =& \sum_{t=1}^z\eoc[(r^*_t\odot\mathcal{E}_t)\backslash((\cup_{j=1}^{t-1} r^*_t\odot\mathcal{E}_t) \cup(\cup_{j=1}^i r_j\odot\mathcal{E}_j))]\cim{Prop.ii}\\
    \leq& \sum_{t=1}^z\eoc[(r^*_t\odot\mathcal{E}_t)\backslash(\cup_{j=1}^i r_j\odot\mathcal{E}_j)]\cim{Prop.iii}\\
    =& \sum_{t=1}^z r^*_t\eoc[\mathcal{E}_t\backslash(\cup_{j=1}^i r_j\odot\mathcal{E}_j)]\cim{Prop.iv}\\
    =& \sum_{t=1}^z r^*_t|\mathcal{Q}_t|\frac{\eoc[\mathcal{E}_t\backslash(\cup_{j=1}^i r_j\odot\mathcal{E}_j)]}{|\mathcal{Q}_t|}\\
    \leq& \sum_{t=1}^z r^*_t|\mathcal{Q}_t|\frac{\eoc[\mathcal{E}_{i+1}\backslash(\cup_{j=1}^i r_j\odot\mathcal{E}_j)]}{|\mathcal{Q}_{i+1}|}\cim{Alg. \ref{algo:sampler}, Line \ref{line:greedy_pick}}\\
    =& \;\beta \cdot \frac{r_{i+1}\eoc[\mathcal{E}_{i+1}\backslash(\cup_{j=1}^i r_j\odot\mathcal{E}_j)]}{r_{i+1}|\mathcal{Q}_{i+1}|}\cim{Def. of $\beta$}\\
    =&\;\frac{\beta}{r_{i+1}|Q_{i+1}|}\eoc[(r_{i+1}\odot\mathcal{E}_{i+1})\backslash(\cup_{j=1}^i r_j\odot\mathcal{E}_j)]\cim{Prop.iv}\\
    =&\;\frac{\beta}{r_{i+1}|Q_{i+1}|}(q_{i+1}-q_{i})\cim{Thm. \ref{theorem:sampler} setup}
\end{align*}
Therefore, $\frac{q^*-q_{i+1}}{q^*-q_{i}}\leq 1-\frac{r_{i+1}|Q_{i+1}|}{\beta}$
\end{proof}
}\vspace{-2mm}
\noindent Now, according to our budget constraint we have
\begin{align*}
    \sum_{n,k} (1-\frac{r_{n,k}|Q_{n,k}|}{\beta})=z-1
\end{align*}
$z$ is the total number of $(n,k)$ pairs where $r_{n,k}>0$, which is a constant. Finally, we have $$\frac{q^*-q}{q^*} =\prod_{i=0}^{z-1}\frac{q^*-q_{i+1}}{q^*-q_{i}}\leq \prod_{n,k}(1-\frac{r_{n,k}|Q_{n,k}|}{\beta})\leq (1-\frac{1}{z})^z< \frac{1}{e}$$
Therefore $q>(1-\frac{1}{e})q*$.

\rv{\subsection{Comparing Our Clique Sampler and Standard Submodular Optimization}\label{subsec:standard_submodular}
\ref{subsec:proof_sampler} suggests two distinctions between our clique sampler and the standard greedy algorithm for submodular optimization.
\begin{itemize}[leftmargin=3mm]
    \item The standard greedy algorithm runs deterministically on a set function whose form is already known. In comparison, our clique sampler runs on a function defined over Random Finite Sets (RFS) whose form can only be statistically estimated from the data. 
    \item  The standard submodular optimization problem forbids picking a set fractionally. Our problem allows fractional sampling from an RFS (\textit{i.e.} $r_{n,k}\in[0,1]$). 
\end{itemize}
As a result, we saw in \ref{subsec:proof_sampler} it is much harder to prove the optimality of our clique sampler than to prove for the greedy algorithm for Standard Submodular Optimization.}

\vspace{-1mm}
\section{Count features}\label{subsec:more_count}
We define a target clique $C=\{v_1, v_2, ..., v_{|C|}\}$. The 8 features are:
{\small
\begin{enumerate}[leftmargin=3mm]
    \item size of the clique: $|C|$;
    \item avg. node degree: $\frac{1}{|C|}\sum_{v\in C}d(v)$;
    \item avg. node degree (recursive): $\frac{1}{|C|}\sum_{v\in C}\frac{1}{|\mathcal{N}(v)|}\sum_{v'\in\mathcal{N}(v)}d(v')$;
    \item avg. node degree \textit{w.r.t.} max cliques: $\frac{1}{|C|}\sum_{v\in C}|\{M\in \mathcal{M}|v\in M\}|$;
    \item avg. edge degree \textit{w.r.t.} max cliques:$\frac{1}{|C|}\sum_{v_1,v_2\in C}|\{M\in \small{\mathsmaller{\mathcal{M}}}|v_1, v_2\in M\}|$;
    \item binarized ``edge degree'' (\textit{w.r.t.} max cliques):$\prod_{v_1,v_2\in C} \mathbbm{1}_{[e]}$, where $e = \sum_{v_1,v_2\in C}|\{M\in \small{\mathsmaller{\mathcal{M}}}|v_1, v_2\in M\}| > 1$;
    \item avg. clustering coefficient: $\frac{1}{|C|}\sum_{v\in C}cc(v)$, where $cc$ is the clustering coefficient of node $v$ in the projection;
    \item avg. size of encompassing maximal cliques: $\frac{1}{|\mathcal{M}^C|}\sum_{M\in \mathcal{M}^C}|M|$, where $\mathcal{M}^C=\{M\in \mathcal{M}|C\subseteq M\}$;
\end{enumerate}
}\vspace{-1mm}
\noindent Notice that avg. clustering coefficient is essentially \rv{a normalized count of the edges between direct neighbors}.

\begin{figure}[t]
    \centering
    \includegraphics[scale=0.42]{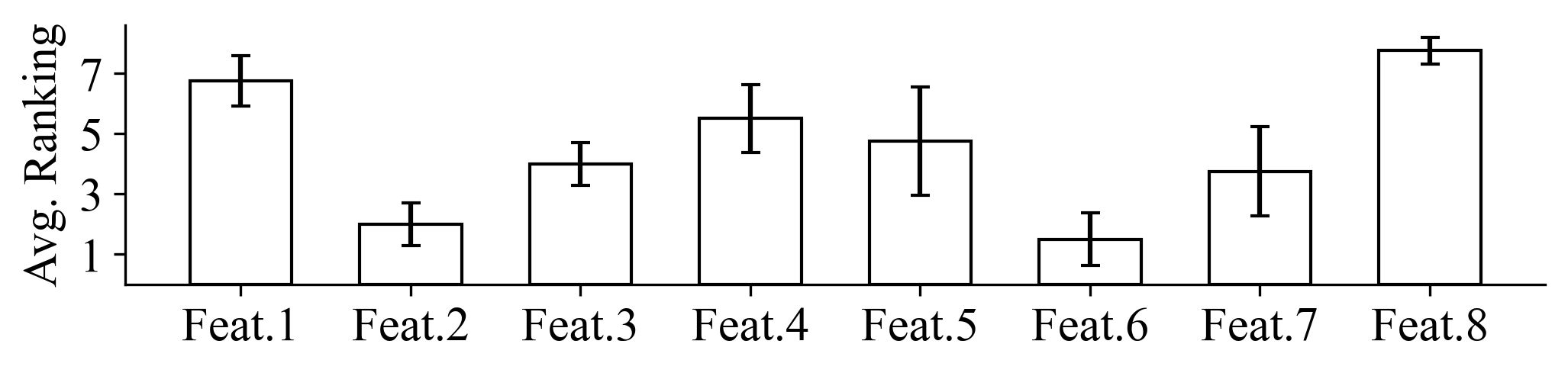}\vspace{-4mm}
    \caption{\rv{Average rankings of the 8 count features. More imporant features have smaller ranking numbers.}}\vspace{-5mm}
    \label{fig:featurerank}
\end{figure}

\rv{\xhdr{Feature Rankings} We study the relative importance of the 8 features by an ablation study. For each dataset, we ablate the 8 features one at a time, record the performance drops, and use those values to rank the 8 features. We repeat this for all datasets, obtaining the 8 features' average rankings, shown in Fig.\ref{fig:featurerank}. More imporant features have smaller ranking numbers. The most important feature is \textit{binarized edge degree}. It is an indicator of ``whether each edge in the target clique has been covered by at least two maximal cliques''.}\vspace{-1mm}

\vspace{-1mm}
\section{Experiments}\label{sec:more_exp}
\subsection{Setup}
\xhdr{Machine Specs} All experiments including model training are run on Intel Xeon Gold 6254 CPU @ 3.15GHz with 1.6TB Memory.

\vspace{-1.7mm}
\xhdr{Datasets} The first 4 datasets in Tab.\ref{tab:dataset} are from \cite{benson2018simplicial}; the rest are from  \cite{young2021hypergraph}. All source links and data are in our supplementary materials.

\vspace{-1.7mm}
\xhdr{Tuning $\beta$} We found the best $\beta$ by training our model on $90\%$ training data and evaluated on the rest $10\%$ training data. The best values are reported in our code instructions.

\vspace{-1.7mm}
\xhdr{Baseline Selection and Adaptation}
For \textit{community detection}, the criteria are: 1. community number must be automatically found; 2. the output is overlapping communities. Based on them, we choose the most representative two. We tested Demon and found it always work best with min community size $=1$ and $\epsilon=1$. To adapt CFinder we search the best $k$ between $[0.1, 0.5]$ quantile of hyperedge sizes on $\mathcal{H}_0$. For \textit{hyperedge prediction}, we ask that they cannot rely on hypergraphs for prediction, and can only use the projection. Based on that we use the two recent SOTAs, \cite{zhang2018beyond,zhang2019hyper}. We use their default hyperparameters for training. For Bayesian-MDL we use its official library in graph-tools with default hyperparameters. We implemented the best heuristic in \cite{conte2016clique} for clique covering.




\subsection{Task Extension: Using Edge Multiplicities}\label{subsec:multiedge}

Throughout this work, we do not assume that the projected graph has
edge multiplicities. Relying on edge multiplicities addresses a
simpler version of the problem which might limit its applicability.
That said, some applications may come with edge multiplicity information,
and it is important to understand what is possible in this more tractable case.
Here we provide an effective unsupervised method as a foundation
for further work.

From Sec. \ref{sec:prelim}, the multiplicity of an edge $(u, v)$
is the number of hyperedges containing both $u$ and $v$.
It is not hard to show that knowledge of the edge multiplicities 
does not suffice to allow perfect reconstruction, and so we still 
must choose from among a set of available cliques to form hyperedges.
In doing this with multiplicity information, we need to ensure that
the cliques we select add up to the given edges multiplicities.
We do this by repeatedly finding maximal cliques, removing them,
and reducing the multiplicities of their edges by 1.
We find that an effective heuristic is to select maximal cliques
that have large size and small average edge multiplicities 
(combining these for example using a weighted sum).

Table \ref{tab:performance_multiedge} gives the performance 
on the datasets we study. We can see that with edge
multiplicities our unsupervised baseline outperforms all the methods
not using edge multiplicities on most datasets, showing the power
of this additional information.
The performance, however, is still far from perfect, 
and we leave the study of this interesting extension to future work.

\begin{table}[H]
\small\begin{tabular}{cccc}
\hline
\textbf{DBLP}       & \textbf{Enron}       & \textbf{P.School}  & \textbf{H.School} \\
82.75 (+1.56)       & 19.79 (+3.77)        & 10.46 (-32.60)     & 19.30 (-35.56)    \\ \hline
\textbf{Foursquare} & \textbf{Hosts-Virus} & \textbf{Directors} & \textbf{Crimes}   \\
83.91 (+10.35)      & 67.86 (+19.01)       & 100.0 (+0.00)      & 80.47 (+1.20)      \\ \hline
\end{tabular}
\caption{\small{\yb{Performance of the proposed baseline using available edge multiplicities. In parenthesis reported the increment against the best-performed method not using edge multiplicities (cr. Table \ref{tab:results}).}}}
\label{tab:performance_multiedge}\vspace{-8mm}
\end{table}

\subsection{Storage Comparison}

\begin{figure}[H]
    \centering
    \includegraphics[scale=0.25]{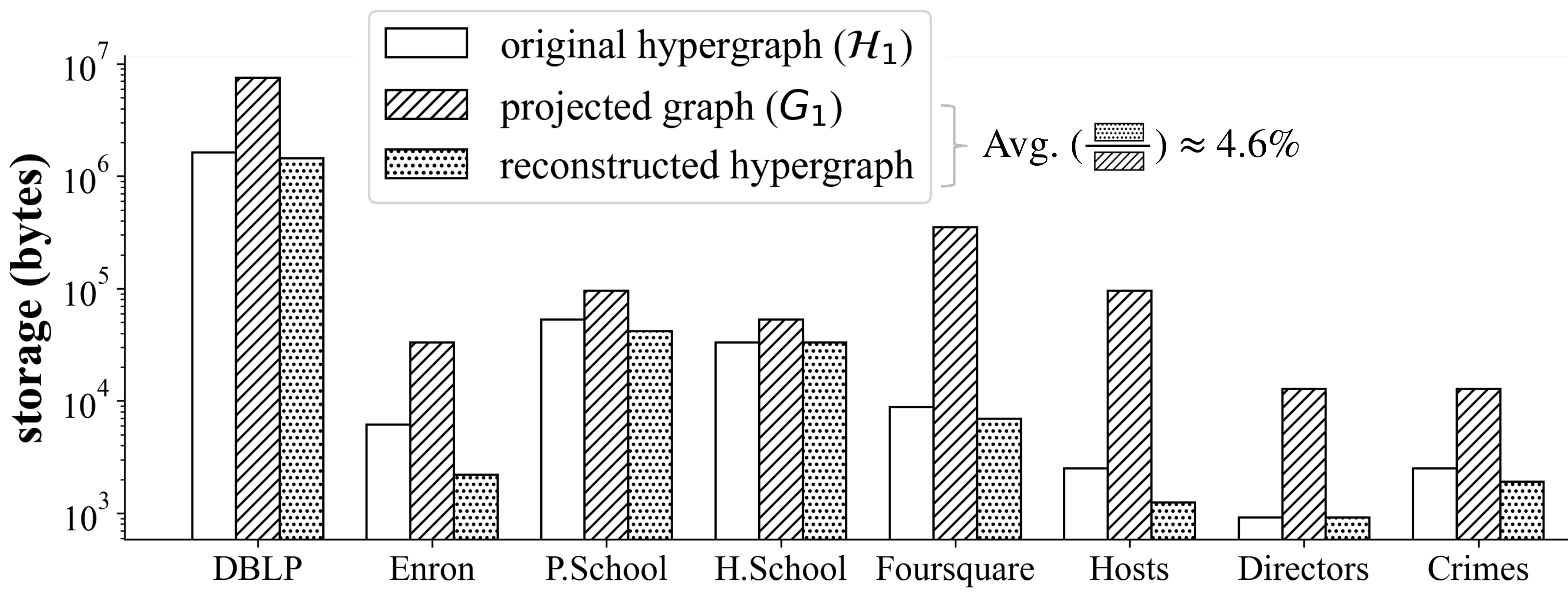}\vspace{-4mm}
    \caption{\small {\yb{Comparing the storage of original hypergraphs, projected graphs and reconstructed hypergraphs. Each hypergraph/graph is stored as a nested array with \texttt{int64} node indexes. Over the 8 datasets on average, a reconstructed hypergraph takes only $4.6\%$ the storage of a projected graph.}}}
    \label{fig:vis_storage}
    \vspace{-4mm}
\end{figure}

As mentioned in Introduction, a side bonus of having a reconstructed hypergraph versus a projected graph is that the former typically requires much less storage space. As a mini-study, we compare the storage of each hypergraph, its projected graph, and its reconstruction generated by \method-count. We use the unified data structure of a nested array to represent the list of edges/hyperedges. Each node is indexed by an int64. Fig.\ref{fig:vis_storage} visualizes the results. We see that the reconstructions take 1 to 2 orders of magnitude less storage space than the projected graphs and are closest to the originals.

\subsection{More Results}

\begin{figure}[H]
    \centering
    \includegraphics[scale=0.35]{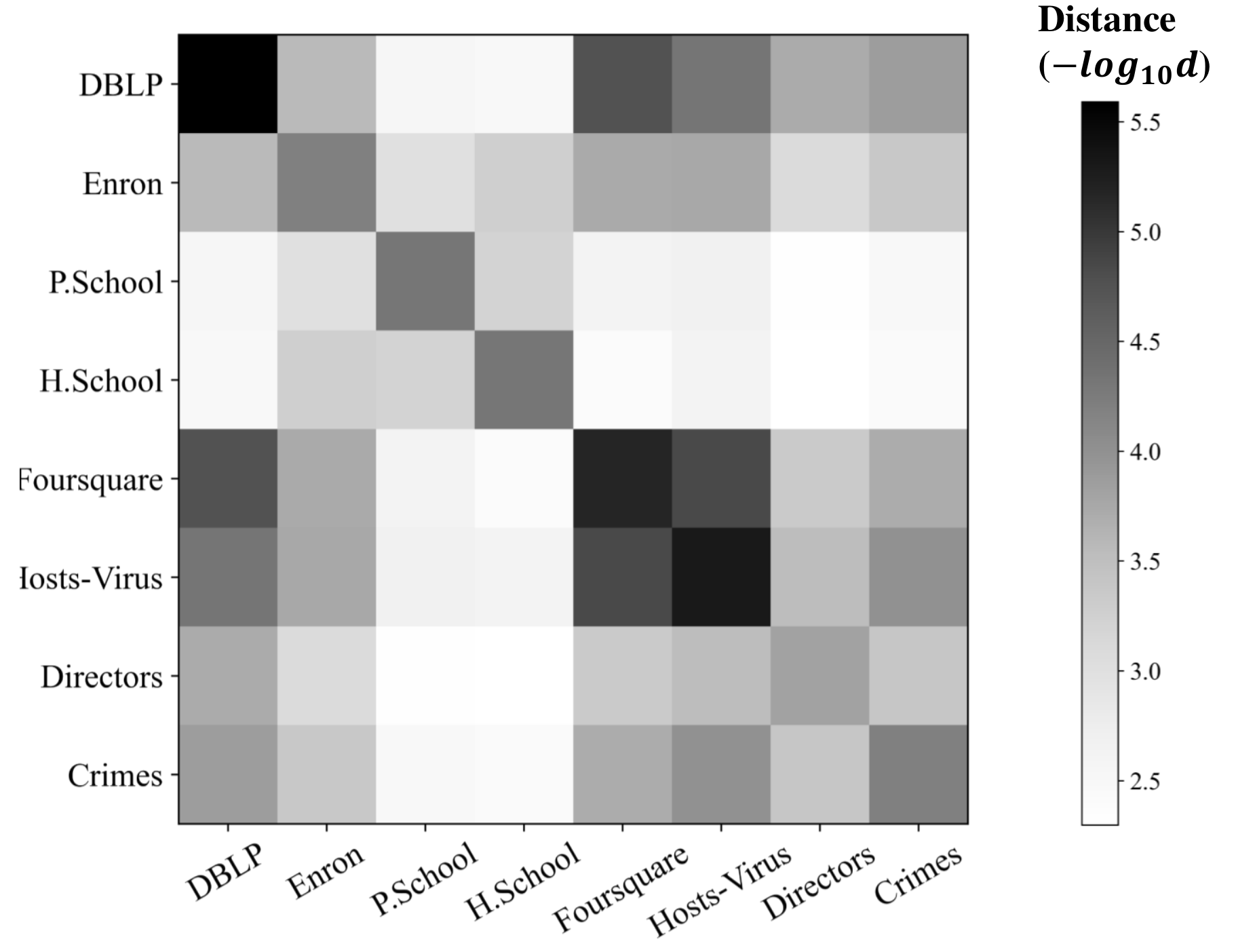}\vspace{-3mm}
    \caption{\rv{Pairwise distance between $\rho(n,k)$ distribution of all datasets using mean squared difference of all cell values (after alignment). The distance matrix obtained is shown above. The diagonal cell is the darkest among its row (or column).}}\label{fig:partitioned_performance}
    \label{fig:distance}\vspace{-2mm}
\end{figure}

\begin{figure}[H]
    \centering\vspace{-1mm}
    \includegraphics[scale=0.38]{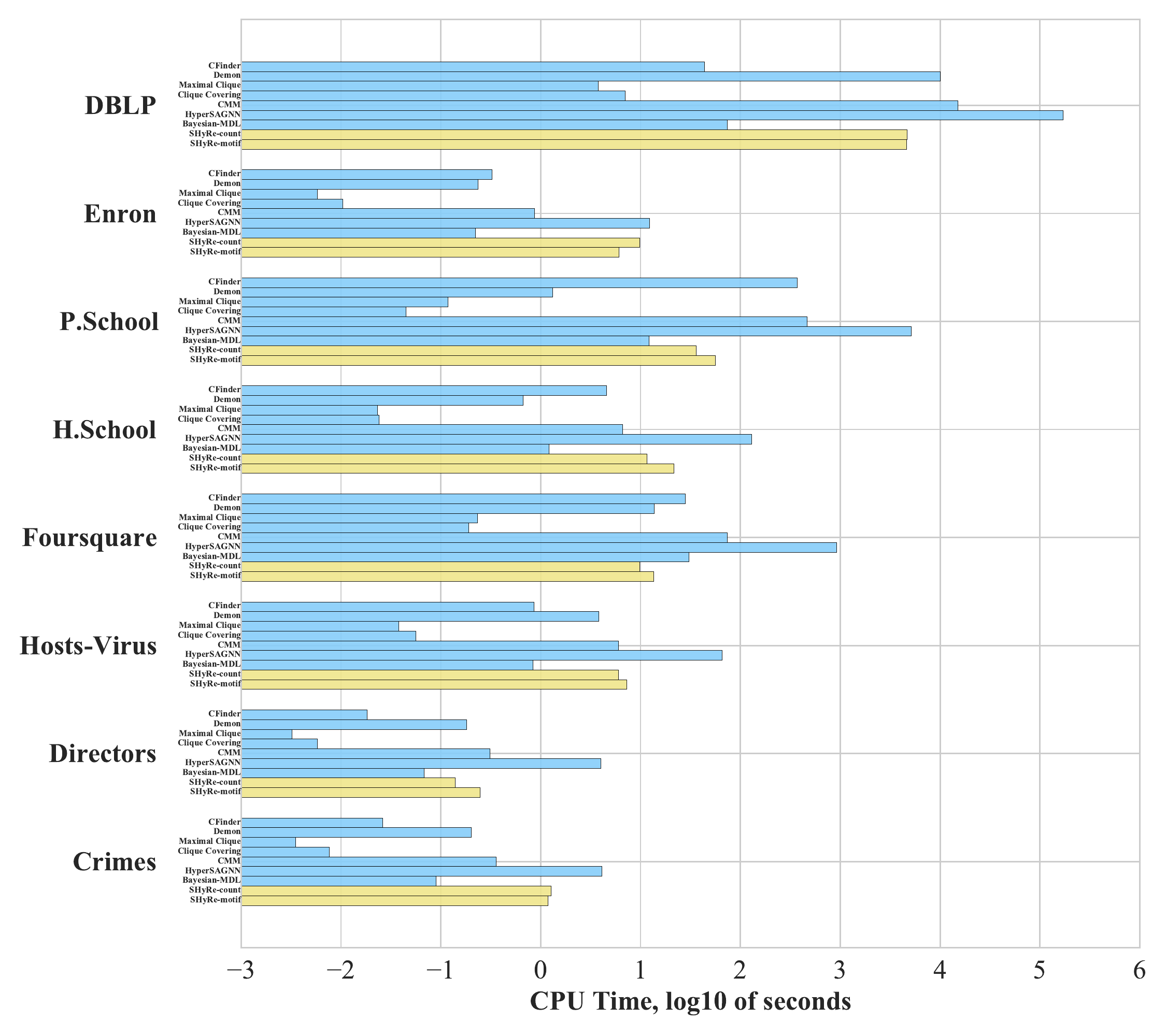}\vspace{-2mm}
    \caption{\rv{CPU time comparison. Notice that Bayesian-MDL's is written in C++, CMM in Matlab, and all other methods in Python.}}
    \label{fig:time}\vspace{-0,5mm}
\end{figure}

\begin{table}[H]\vspace{-1mm}
\rv{\small\begin{tabular}{cccc}
\hline
\textbf{DBLP}       & \textbf{Enron}       & \textbf{P.School}  & \textbf{H.School} \\
$1.0E6$, $(\footnotesize{\gg}6.0E10$)      & $1.0E3$, $6.9E4$    &    $3.5E5$, $2.6E6$      & $6.0E4$, $8.4E5$      \\ \hline
\textbf{Foursquare} & \textbf{Hosts-Virus} & \textbf{Directors} & \textbf{Crimes}   \\
$2.0E4$, $(\gg1.1E12$)      & $6.0E3$, ($\gg2.2E12$)  &    $800$, $4.5E5$      & $1.0E3$, $2.9E5$      \\ \hline
\end{tabular}}
\caption{\small{\rv{Optimal clique sampling number $\beta$ and total number of cliques $|\mathcal{U}|$. ``$\gg$'' appears if the dataset contains too many cliques to be enumerated by our machine in 24 hours, in which case a conservative lower bound is estimated instead.}}}
\label{tab:beta}
\end{table}

\begin{table}[H]
\footnotesize\vspace{-4mm}
\begin{tabular}{lccc}
\hline
\textbf{}      & \textbf{DBLP}  & \textbf{Hosts-Virus} &\textbf{Enron}   \\ \hline
original (\method-motif) & 81.19$\pm$0.02 & 45.16$\pm$0.55 &16.02$\pm$0.35        \\
ablation: ``random''  & 0.17$\pm$0.00 & 0.00$\pm$0.00  &0.54$\pm$0.49        \\
ablation: ``small'' & 1.12$\pm$0.52  & 1.38$\pm$0.70  & 8.57$\pm$0.89       \\ 
ablation: ``head \& tail'' & 27.42$\pm$0.54 & 29.92$\pm$0.54 & 11.99$\pm$0.10             \\ 
\hline
\end{tabular}
\caption{Ablation study: comparing the performance obtained by replacing the clique sampler with simpler heuristics for sampling.}\label{tab:exp_ablation}\vspace{-3mm}
\end{table}



\end{document}